%% file: main.tex
\documentclass[lettersize,journal]{IEEEtran}
\usepackage{amsmath,amsfonts}
\usepackage{algorithmic}
\usepackage{algorithm}
\usepackage{array}
\usepackage[caption=false,font=normalsize,labelfont=sf,textfont=sf]{subfig}
\usepackage{textcomp}
\usepackage{stfloats}
\usepackage{url}
\usepackage{verbatim}
\usepackage{graphicx}
\usepackage{cite}
\usepackage{enumerate}
\usepackage{algorithm}
\usepackage{algorithmic}
\usepackage{hyperref}
\usepackage{booktabs}
\usepackage{multirow}
\usepackage{graphicx}
\usepackage{xcolor}
\usepackage{pifont}
\usepackage{amsmath,amsfonts}
\usepackage{amssymb}
\usepackage{graphicx}
\usepackage{textcomp}
\usepackage{xspace}
\usepackage{lipsum}
\usepackage{makecell}
\usepackage{array}
\usepackage{url}
\usepackage{wasysym}
\usepackage{enumitem}
\usepackage[table,dvipsnames]{xcolor}
\setlist[itemize]{leftmargin=*}
\setlist[enumerate]{leftmargin=*}
\newtheorem{definition}{Definition}
\newcommand{\sys}{$\mathsf{EmbTracker}$\xspace}

\begin{document}

\title{\sys: Traceable Black-box Watermarking for Federated Language Models}

\author{Haodong Zhao, Jinming Hu, Yijie Bai, Tian Dong, Wei Du, Zhuosheng Zhang,\\ Yanjiao Chen,~\IEEEmembership{IEEE Senior Member}, Haojin Zhu,~\IEEEmembership{IEEE Fellow}, Gongshen Liu
\thanks{This work is partially supported by the Joint Funds of the National Natural Science Foundation of China (Grant No.U21B2020), Special Fund for the Action Plan of Shanghai Jiao Tong University's ``Technological Revitalization of Mongolia'' under Subcontract No.2025XYJG0001-01-06, National Natural Science Foundation of China (62406188) and Natural Science Foundation of Shanghai (24ZR1440300). (Corresponding author: Zhuosheng Zhang, Gongshen Liu)}
\thanks{Haodong Zhao, Jinming Hu, Zhuosheng Zhang, Haojin Zhu and Gongshen Liu are with School of Computer Science, Shanghai Jiao Tong University, Shanghai, China. Gongshen Liu is also with Inner Mongolia Research Institute, Shanghai Jiao Tong University
(e-mail: \{zhaohaodong, hujinming, zhangzs, zhu-hj, lgshen\}@sjtu.edu.cn).}
\thanks{Yijie Bai and Wei Du are with Ant Group, China (e-mail: \{baiyijie.byj, xiwei.dw\}@antgroup.com).}
\thanks{Tian Dong is with The University of Hong Kong, China (e-mail: tiandong@hku.hk).}
\thanks{Yanjiao Chen is with the College of Electrical Engineering, Zhejiang University, Hangzhou, China (e-mail: chenyanjiao@zju.edu.cn).}}

\markboth{Journal of \LaTeX\ Class Files,~Vol.~14, No.~8, August~2021}%
{Shell \MakeLowercase{\textit{et al.}}: A Sample Article Using IEEEtran.cls for IEEE Journals}

\IEEEpubid{0000--0000/00\$00.00~\copyright~2021 IEEE}

\maketitle

\begin{abstract}

Federated Language Model (FedLM) allows a collaborative learning without sharing raw data, yet it introduces a critical vulnerability, as every untrustworthy client may leak the received functional model instance. Current watermarking schemes for FedLM often require white-box access and client-side cooperation, providing only group-level proof of ownership rather than individual traceability. We propose \sys, a server-side, traceable black-box watermarking framework specifically designed for FedLMs. \sys achieves black-box verifiability by embedding a backdoor-based watermark detectable through simple API queries. Client-level traceability is realized by injecting unique identity-specific watermarks into the model distributed to each client. In this way, a leaked model can be attributed to a specific culprit, ensuring robustness even against non-cooperative participants.
Extensive experiments on various language and vision-language models demonstrate that \sys achieves robust traceability with verification rates near 100\%, high resilience against removal attacks (fine-tuning, pruning, quantization), and negligible impact on primary task performance (typically within 1-2\%).
\end{abstract}


\input{sections/1introduction}
\input{sections/2preliminaries}
\input{sections/3threat}
\input{sections/4design}
\input{sections/5evaluation}
\input{sections/7relatedworks}

\input{sections/8conclusion}

\bibliographystyle{IEEEtran}
\bibliography{refs}
\vspace{-18mm}
\begin{IEEEbiography}[{\includegraphics[width=1in,height=1.25in,clip,keepaspectratio]{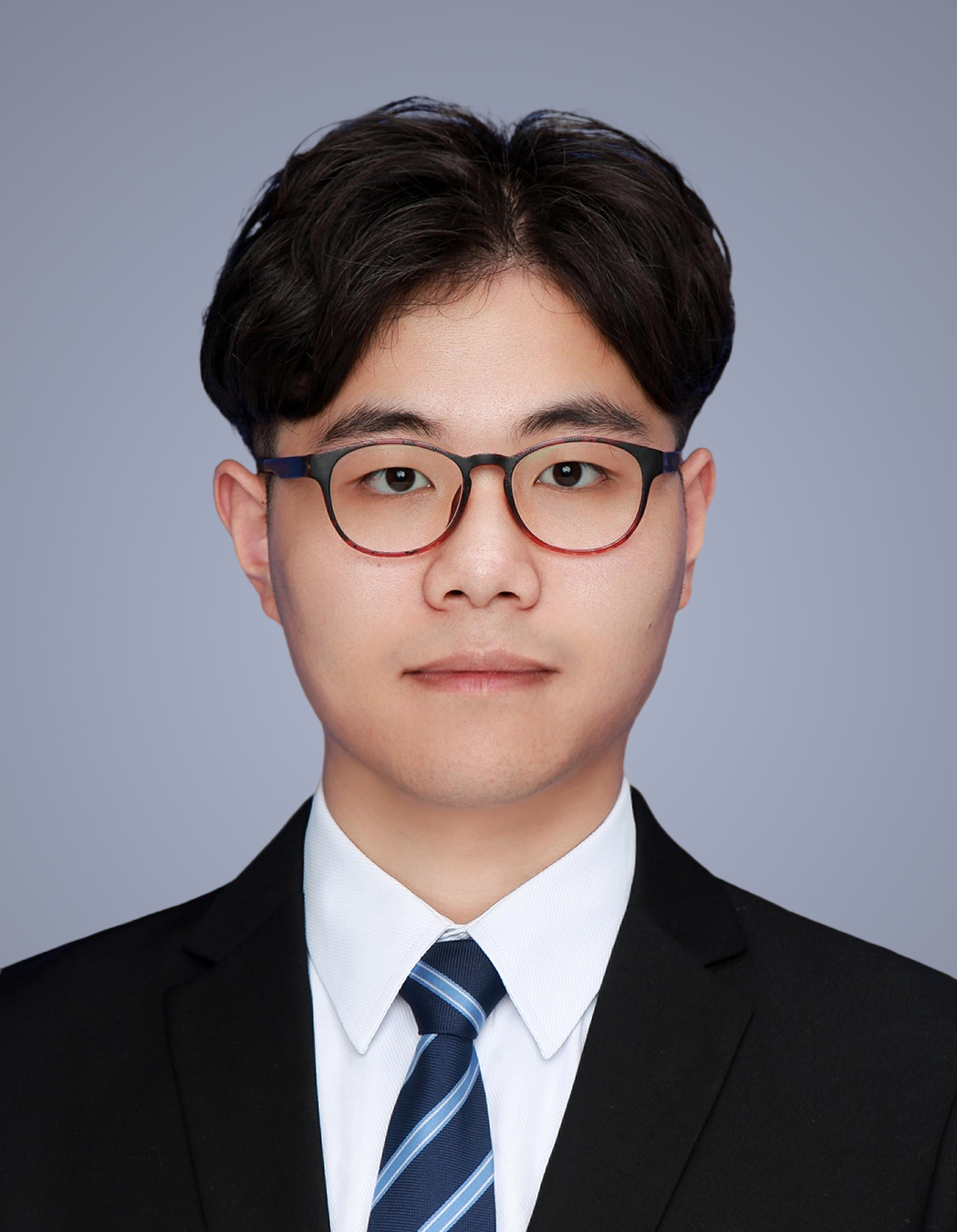}}]{Haodong Zhao} (Student Member, IEEE)
received his bachelor's degree from Shanghai Jiao Tong University (SJTU), in 2021.  He is currently working toward the PhD degree in School of Computer Science,
Shanghai Jiao Tong University. His research interests include LLM-Agent, NLP, Federated Learning, and AI security.
\end{IEEEbiography}
\vspace{-18mm}
\begin{IEEEbiography}[{\includegraphics[width=1in,height=1.25in,clip,keepaspectratio]{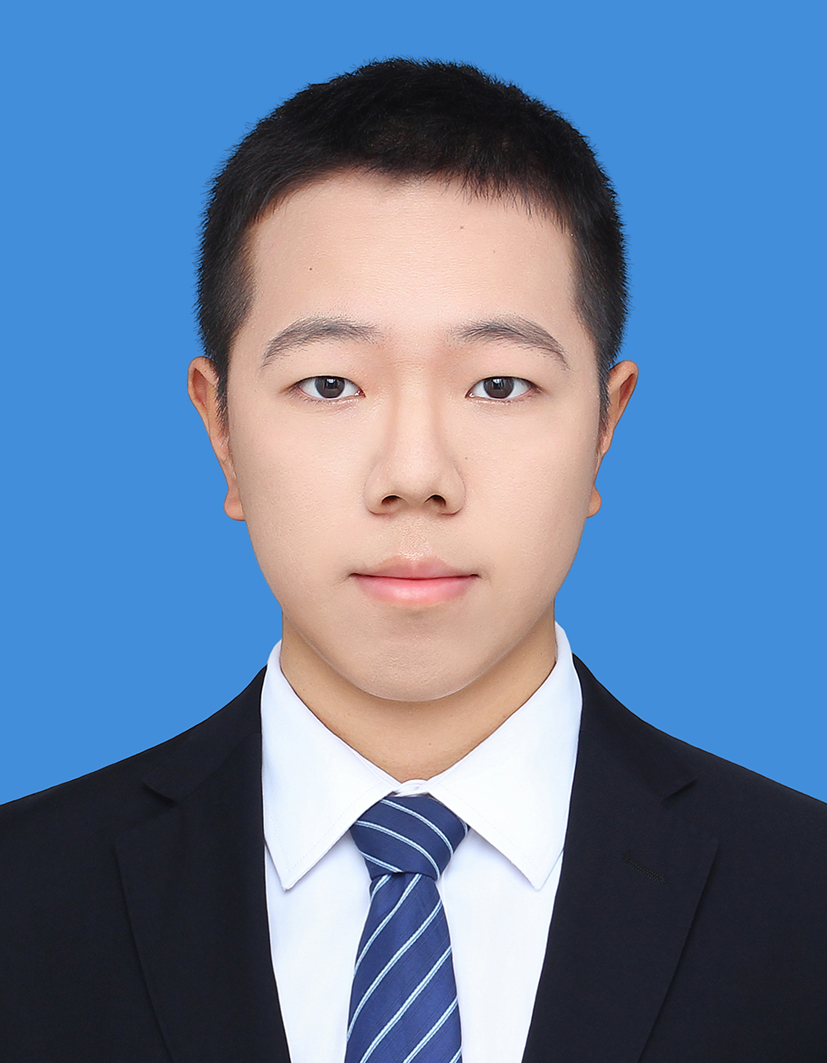}
}]{Jinming Hu} is an undergraduate student in the School of Computer Science, Shanghai Jiao Tong University, expected to receive the B.Eng. degree in 2026. His research interests include AI security and natural language processing.
\end{IEEEbiography}
\vspace{-15mm}
\begin{IEEEbiography}[{\includegraphics[width=1in,height=1.2in,clip,keepaspectratio]{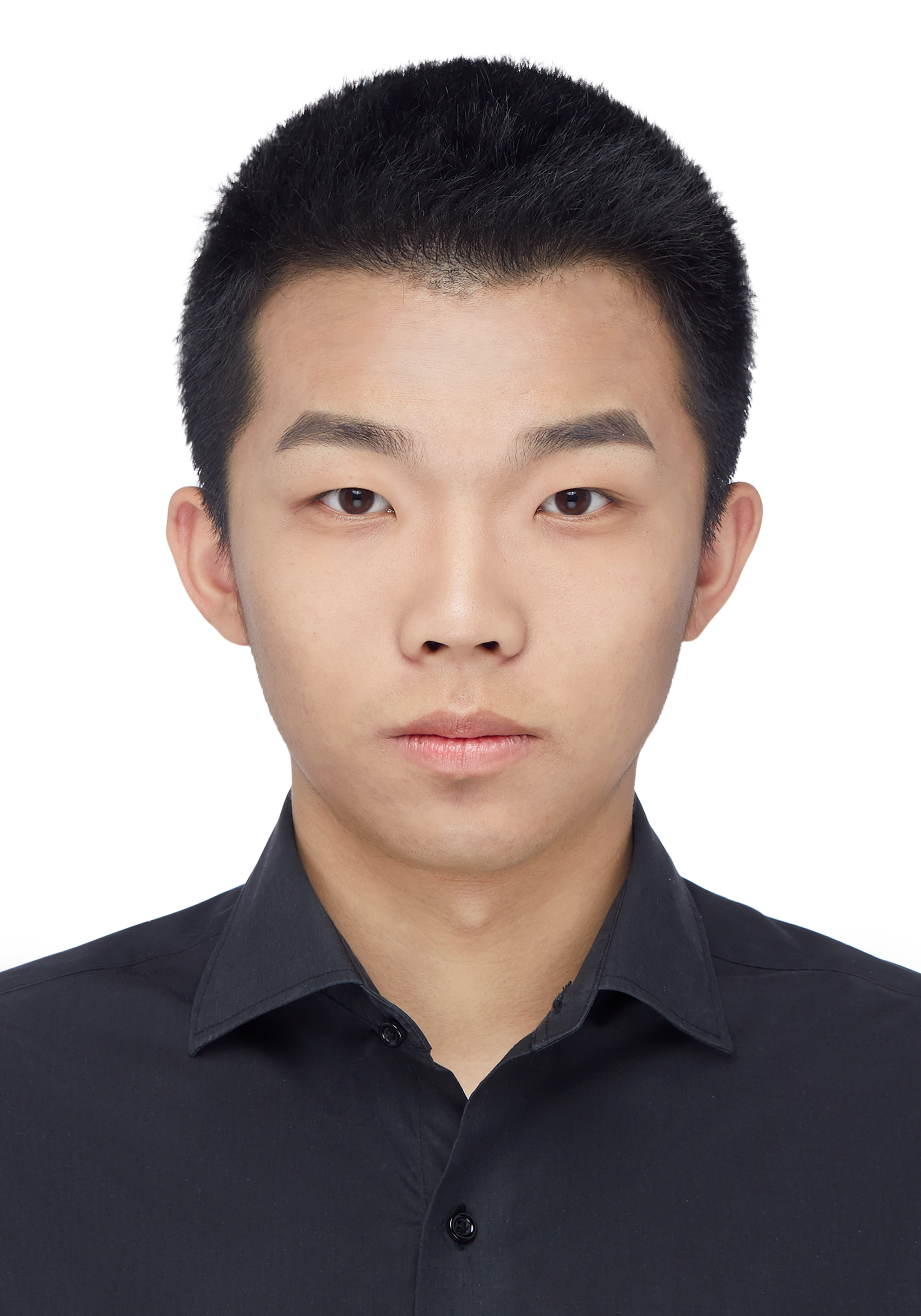}
}]{Yijie Bai} received his B.E. degree from the Department of Automation, Tsinghua University, China, in 2020, and his Ph.D. degree from the College of Electrical Engineering, Zhejiang University, China. He is currently conducting research on large model security at Ant Group. His research interests include machine learning security, data privacy, and federated learning.
\end{IEEEbiography}
\vspace{-15mm}
\begin{IEEEbiography}[{\includegraphics[width=1in,height=1.2in,clip,keepaspectratio]{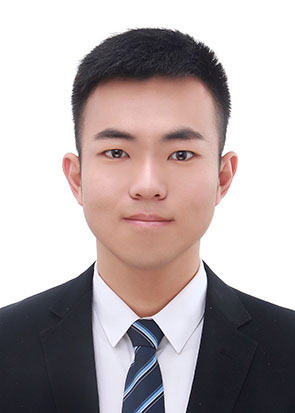}
}]{Tian Dong} (Graduate Student Member, IEEE)
received the B.A., M.E. and Ph.D. degree from Shanghai Jiao Tong
University, China, in 2019, 2022 and 2025.
He is currently a postdoctoral fellow at the University of Hong Kong.
His research
interests include the intersection of security, privacy, and machine learning.
\end{IEEEbiography}
\vspace{-15mm}
\begin{IEEEbiography}[{\includegraphics[width=1in,height=1.1in,clip,keepaspectratio]{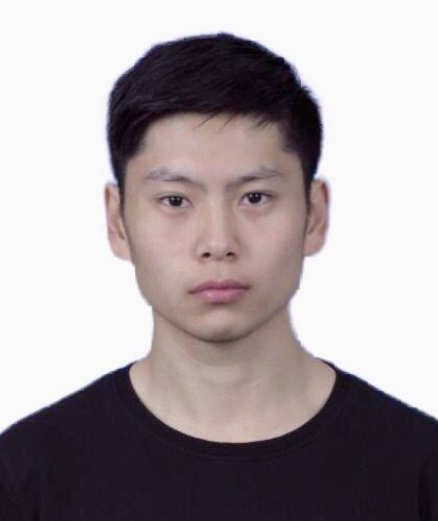}
}]{Wei Du} received the B.S Degree from the School
of Electronic Engineering, XiDian University in 2020, and the Ph.D. Degree with the School of Cyber Science and Engineering, Shanghai Jiao Tong University in 2025. His primary research interests include natural language processing, artificial intelligent security and backdoor attacks.
\end{IEEEbiography}
\vspace{-15mm}
\begin{IEEEbiography}[{\includegraphics[width=1in,height=1.25in,clip,keepaspectratio]{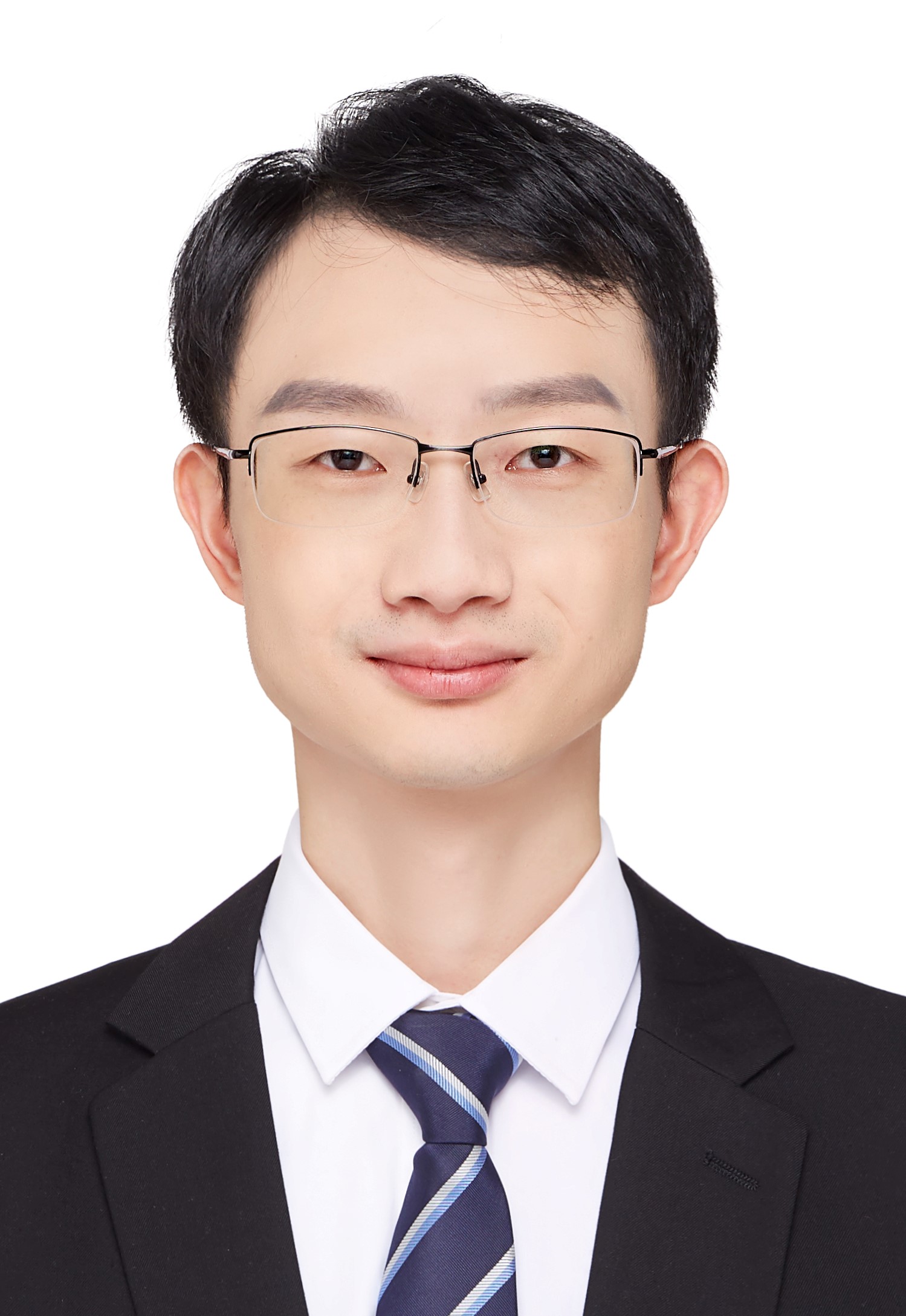}}]{Zhuosheng Zhang} received his Bachelor's degree in internet of things from Wuhan University in 2016, his M.S. degree and his Ph.D. degree in computer science from Shanghai Jiao Tong University in 2020 and 2023. He is currently an assistant professor at Shanghai Jiao Tong University. He was an intern at NICT, Microsoft Research, and Amazon Web Services. His research interests include natural language processing, large language models, and language agents.
\end{IEEEbiography}
\vspace{-14mm}
\begin{IEEEbiography}[{\includegraphics[width=1in,height=1.25in,clip,keepaspectratio]{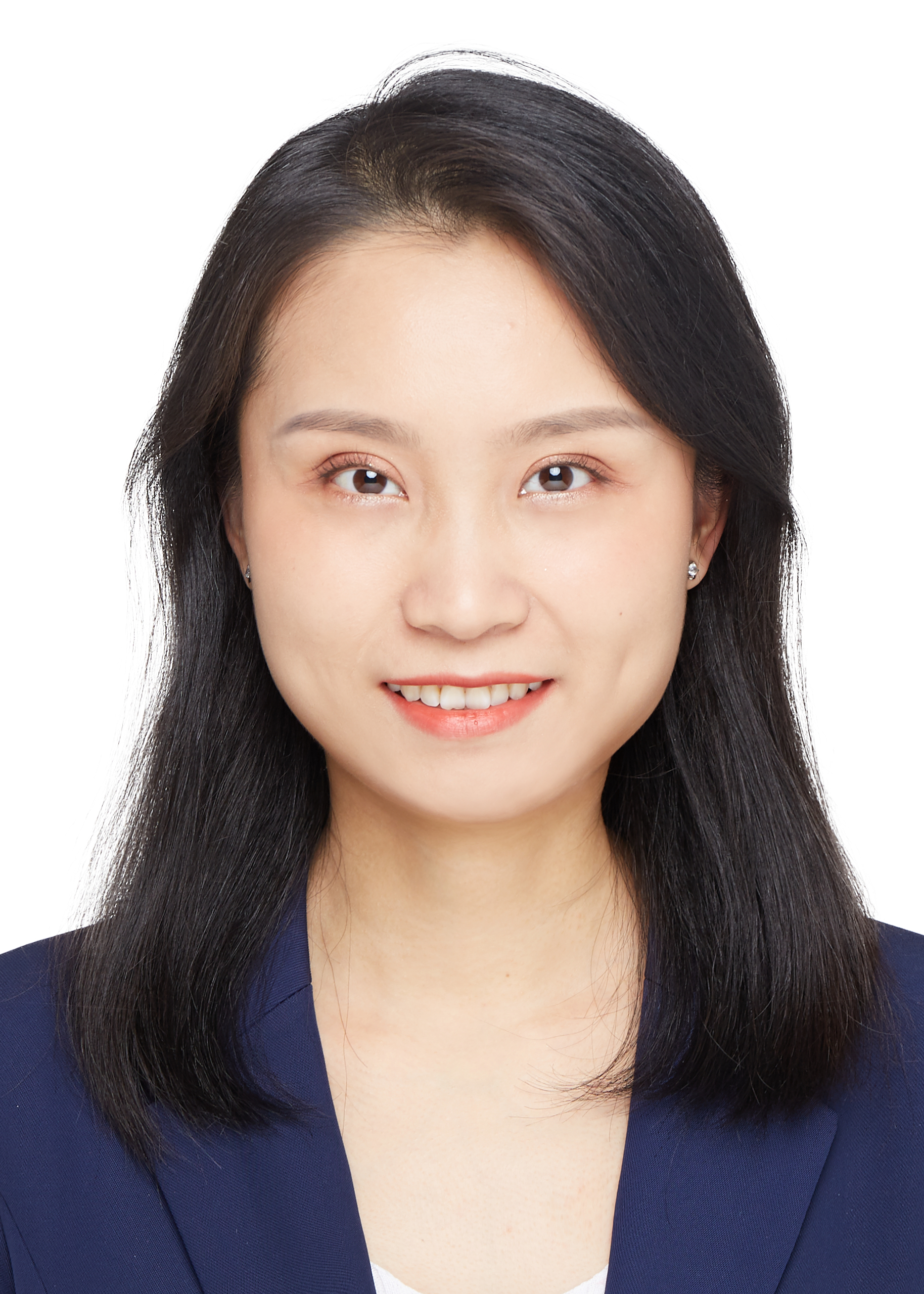}
}]{Yanjiao Chen} (Senior Member, IEEE) received her B.E. degree in Electronic Engineering from Tsinghua University in 2010 and Ph.D. degree in Computer Science and Engineering from Hong Kong University of Science and Technology in 2015. She is currently a Bairen Researcher in Zhejiang University, China. Her research interests include AI security, Smart IoT security, and network security.
\end{IEEEbiography}
\vspace{-12mm}
\begin{IEEEbiography}[{\includegraphics[width=1in,height=1.25in,clip,keepaspectratio]{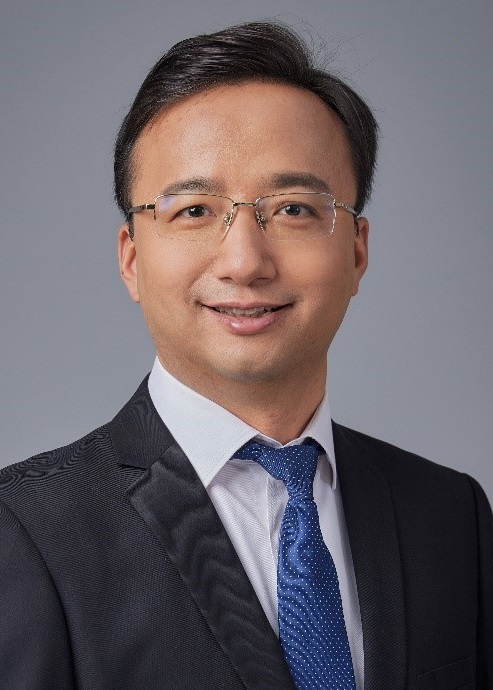}
}]{Haojin Zhu} (Fellow, IEEE) received the BSc degree in computer science from Wuhan University, China, in 2002, the MSc degree in computer science from Shanghai Jiao Tong University, China, in 2005, and the PhD degree in electrical and computer engineering from the University of Waterloo, Canada, in 2009; he is currently a professor with the Computer Science Department, Shanghai Jiao Tong University; he has published more than 160 papers in top-tier conferences including IEEE S\&P, ACM CCS, USENIX Security, and NDSS; he has received a number of awards including the SIGSOFT Distinguished Paper Award of ESEC/FSE (2023), ACM CCS Best Paper Runner-Ups Award (2021), and USENIX Security Distinguished Paper Award (2024); his current research interests include network security and privacy enhancing technologies.
\end{IEEEbiography}
\vspace{-10mm}

\begin{IEEEbiography}[{\includegraphics[width=1in,height=1.25in,clip,keepaspectratio]{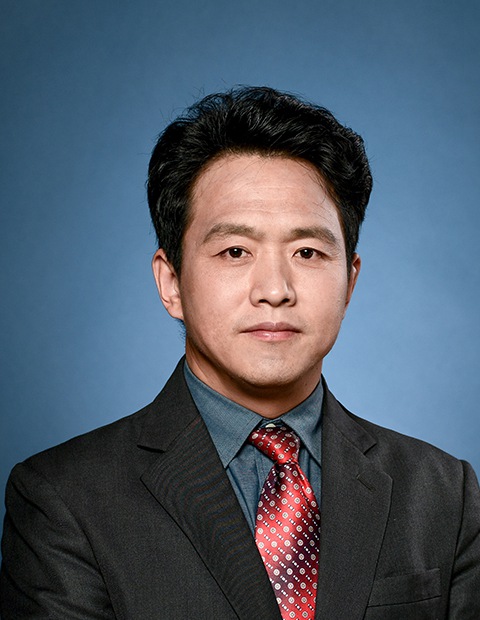}}]{Gongshen Liu} received his Ph.D. degree in Department of Computer Science from Shanghai Jiao Tong University. He is currently a professor with the School of Electronic Information and Electrical Engineering, Shanghai Jiao Tong University. His research interests cover Natural Language Processing, Machine Learning and Artificial Intelligent Security. 
\end{IEEEbiography}

\vfill

\end{document}

%% file: sections/1introduction.tex
\section{Introduction}
\IEEEPARstart{F}{ederated} language model (FedLM) training has become a practical way to fine-tune language models (LMs) across distributed data silos (e.g., enterprises, institutions, and user devices) while keeping raw data local~\cite{mcmahan2017communication,ye2024fedllm}. In a typical FedLM pipeline, a server repeatedly distributes a global LM to clients, clients perform local updates on private text, and the server aggregates updates to improve the shared model. While this setting improves data governance, it also amplifies IP leakage risk: any participating client can obtain a high-value model snapshot during training and redistribute it without authorization~\cite{yu2023leaked}.

\begin{figure}[!t]
    \centering
\includegraphics[width=1.0\columnwidth]{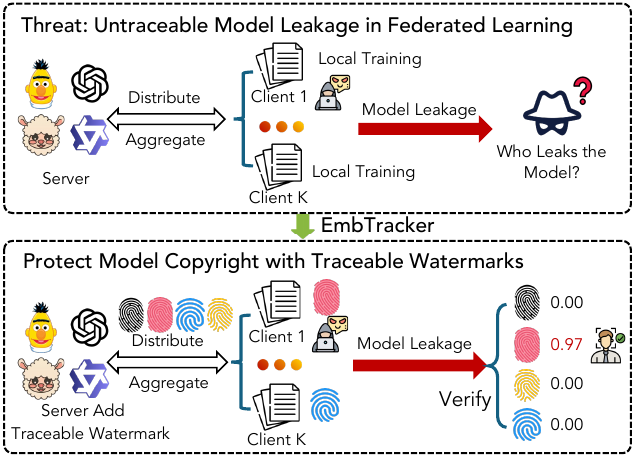}
\vspace{-18pt}
    \caption{Illustration of the risk of client model leakage in federated language model training. Since all clients in FL can obtain the same global model, traditional watermarks cannot distinguish the source of the leak. \sys creates a unique watermark for each client through the server, which can accurately track the model leaker.}
    \label{fig:threat}
    \vskip -0.2in
\end{figure}

Digital watermarking is a natural direction for IP protection~\cite{UchidaNSS17}, but FedLM imposes requirements that existing federated watermarking methods do not simultaneously satisfy. In particular, when a suspicious LM is found in the wild, the defender often has only black-box access (e.g., an API endpoint) rather than model parameters, and the defender must answer not only ``is this our model?''~\cite{tekgul2021waffle} but also ``which client leaked it?''~\cite{shao2022fedtracker,xu2025traceable} as shown in Fig.~\ref{fig:threat}. However, most FL watermarking solutions either (i) embed a single watermark shared by all clients (supporting only group-level ownership)~\cite{yu2023leaked}, (ii) rely on white-box parameter inspection~\cite{li2022fedipr,liang2023fedcip,yang2023fedsov,xu2024robwe,shao2022fedtracker}, or (iii) require client-side participation for embedding/verification~\cite{li2022fedipr,liang2023fedcip,yang2023fedsov,xu2024robwe,nie2024persistverify}. These assumptions are fragile in adversarial federations and misaligned with real-world LM deployment. Moreover, FedLMs commonly use parameter-efficient fine tuning (PEFT) methods~\cite{han2024parameter}, and LMs include generation behaviors that are not covered by specialized watermark designs for vision or simple classification models.
\IEEEpubidadjcol

\begin{table}[]
\centering
\renewcommand{\arraystretch}{0.9}
\caption{Comparison of representative FL watermarking methods. \CIRCLE~and \Circle~denotes black-box and white-box verification, respectively. ``S'' and ``C'' indicate watermarking by server and client. For the method that appears twice, it means that there are two watermarking schemes.}
\label{tab:comp}
\resizebox{0.99\linewidth}{!}{
\begin{tabular}{lcccc}
\toprule
\textbf{Method}    & \textbf{Domain} & \textbf{Verification} & \textbf{Injector} & \textbf{Traceability} \\ \midrule
WAFFLE~\cite{tekgul2021waffle}       & CV   & \CIRCLE    & S         & \ding{55} \\
FedIPR~\cite{li2022fedipr}           & CV   & \CIRCLE    & C       & \ding{51} \\
FedIPR~\cite{li2022fedipr}           & CV   &  \Circle    & C       & \ding{51} \\
FedCIP~\cite{liang2023fedcip}         & CV   & \Circle    & C         & \ding{51} \\
FedSOV~\cite{yang2023fedsov} & CV, NLP & \Circle & C & \ding{51} \\
RobWE~\cite{xu2024robwe}             & CV &  \Circle & C      & \ding{51} \\
FedTracker~\cite{shao2022fedtracker} & CV   &  \CIRCLE    & S         & \ding{55} \\
FedTracker~\cite{shao2022fedtracker} & CV   & \Circle  & S         & \ding{51} \\
PersistVerify~\cite{nie2024persistverify} & CV & \CIRCLE & C         & \ding{55} \\
TraMark~\cite{xu2025traceable}       & CV   & \CIRCLE    & S         & \ding{51} \\
\textbf{\sys (Ours)}                          & NLP  & \CIRCLE    & S         & \ding{51} \\
\bottomrule
\end{tabular}}
\vskip -0.15in
\end{table}
Table~\ref{tab:comp} provides a comparision, these limitations suggest that new methods are needed to overcome challenges:

\begin{itemize}
\setlength{\itemsep}{10pt}
    \item  \textit{Challenge 1: Black-box verifiability for deployed LMs.\looseness=-1
}
\end{itemize}

In realistic leakage threats, the model owner typically cannot access internal weights and can only query the suspicious model. A practical FedLM watermark must therefore be verifiable through black-box queries, robust across LM tasks, and compatible with federated PEFT workflows, without requiring client-side changes that could be refused.

\begin{itemize}
\setlength{\itemsep}{10pt}
    \item  \textit{Challenge 2: Client traceability w/o prohibitive overhead.\looseness=-1
}
\end{itemize}

Traceability requires non-colliding, identity-specific watermarks, and models distributed to different clients should respond differently to verification queries so the leaker can be uniquely identified. Achieving this at scale is difficult because naive approaches demand per-client retraining or complicated protocols, and continuous updates from training potentially weaken fragile watermarks. A deployable solution should provide uniqueness per-client with negligible extra overhead and remain stable and robust.

To address these challenges, we propose \sys, a server-side framework for traceable black-box watermarking in FedLMs. \sys uses an embedding-based backdoor watermark that is (i) black-box verifiable through trigger queries and (ii) client-traceable by issuing a uniquely watermarked model to each client. The key insight is that the word embedding space provides a high-capacity, low-interference carrier for watermark signals in modern LMs. Modifying several embeddings is difficult to notice, yet effective in enforcing trigger-response behavior. In practice, the server first learns a universal watermark embedding by updating only the embedding vector of the universal trigger (one-time cost). It then produces client-specific watermarks by efficiently mapping each client’s identity to a distinct trigger and replacing the corresponding trigger embedding before distribution, avoiding per-client retraining and remaining compatible with common PEFT methods such as LoRA~\cite{hu2022lora} and prefix tuning~\cite{li2021prefix}. When a suspicious model is discovered, the defender can perform black-box tracing by querying with each client’s trigger set and attributing the leak based on verification.

Our main contributions are summarized as follows:

$\bullet$ We propose \sys, the first FedLM-tailored server-side traceable black-box watermarking framework, allowing client-level attribution of model leakage without any client-side modifications.

$\bullet$ We introduce an efficient embedding-based watermark injection method that embeds identity-specific watermarks in the word embedding space with negligible overhead and wide compatibility across PEFT strategies.

$\bullet$ Extensive experiments on classification and generation tasks demonstrate that \sys achieves strong traceability, high robustness to removal attacks, and minimal impact on task accuracy, outperforming existing traceable methods.

%% file: sections/2preliminaries.tex
\section{Preliminaries}
\subsection{Federated Learning for Language Models}
In a classical FL setting, consider a single server (aggregator) $S$ and $K$ clients, denoted as $C=\{c_k \mid k \in [1,K]\}$. In many scenarios, the server may also participate as one of the clients. Each client $c_k$ has a private data set $\mathcal{D}_{k} = \{(x_k, y_k)\}$ comprising $n_k$ samples.
During each communication round $r$, the server $S$ distributes the current global model $M^r$ to all clients and subsequently collects the $k$-th local model update $M^r_k$ (or the corresponding gradients), which share the same architecture as the global model. 

Each round of the FL process can be summarized:
\begin{enumerate}
    \item The global model at round $r$, denoted as $M^r$, is distributed by the central server $S$ to all clients.
    \item Upon receipt of $M^r$, each client $c_k$ undertakes local optimization utilizing its private dataset and loss function $L_k$. The local model update is computed using
        $M^r_k = M^r - \eta \cdot \frac{\partial L_k}{\partial M^r}$,    
    where $\eta$ represents the learning rate. Subsequently, the locally updated model parameters $M^r_k$ are transmitted back to the server.
    \item The server aggregates the received updates via protocol like FedAvg~\cite{mcmahan2017communication}, to synthesize the updated global model for the next round, $M^{r+1} = \sum_{k=1}^K \frac{n_k}{n} M^r_k$,    
    where $n = |\mathcal{D}| = \sum_{k=1}^K n_k$ denotes the aggregate number of training samples across all clients, and $\mathcal{D} \triangleq \bigcup_{k=1}^K \mathcal{D}_k$ signifies the union of individual client datasets.
\end{enumerate}

As LMs have demonstrated remarkable capabilities through robust and versatile architectures, FL has gained significant attention as a scalable, privacy-preserving approach for training LMs on data silos~\cite{zhao2023fedprompt,ye2024fedllm,wu2025survey,bian2025survey,zhao2026revisiting,zhao2026protegofed}. However, the conventional approach, known as full fine-tuning, which updates all model parameters on client devices, is usually impractical for FL of LMs due to the substantial parameter transmission overhead across devices.
To address this challenge, PEFT methods~\cite{han2024parameter} such as LoRA~\cite{hu2022lora} and prefix tuning~\cite{li2021prefix} have emerged as effective solutions. By January 20, 2024, the number of adapters in huggingface had exceeded 10,000~\cite{sun2025peftguard}. Using PEFT techniques in FL allows clients to update only a small subset of model parameters or introduce lightweight trainable modules, enabling efficient local adaptation of LLMs while keeping the majority of the model fixed. Based on this, only a small set of trainable parameters is exchanged with the server for aggregation. As a result, FL combined with PEFT unlocks the potential to collaboratively train powerful LLMs in diverse and resource-constrained settings, and PEFT-based methods, such as LoRA, have become the main solution~\cite{wu2025survey,bian2025survey}.\looseness=-1

\subsection{Model Watermarking}

IP protection for DNNs has recently garnered significant attention, particularly with the rapid advancement of LLMs, which require substantial computational resources, human expertise, and proprietary organizational knowledge. Technically, model watermarking schemes typically consist of two primary phases: watermark injection and watermark verification. Depending on the verification phase, watermarking schemes are commonly classified into two categories: white-box watermarking and black-box watermarking.\looseness=-1

\textbf{White-box Watermarking.}
In white-box watermarking schemes, models are distinguished by unique marks embedded in their structure or parameters~\cite{chen2021you,fan2019rethinking}. During the injection phase, identity-representative signature messages are incorporated into the model either through additional training or by directly modifying the model parameters. White-box watermarking assumes that the verifier has complete access to the suspect model in verification. This enables the verifier to inspect the model's structure and parameters, extract the embedded secret message, and compare it with the owner's reference. However, this assumption is often unrealistic in practical scenarios, where models are typically accessed only via black-box interfaces. As a result, the applicability of white-box watermarks is inherently limited.

\textbf{Black-box Watermarking.}
Black-box watermarking schemes relax the requirements by assuming that the verifier can only observe the output of the suspect model. In the context of deep neural networks, backdoor attacks are well-aligned with black-box verification, and thus black-box watermarks are often embedded using backdoor-based techniques. Typically, a set of trigger inputs (e.g., task-irrelevant images or rare words) is designated as the watermark, with special labels assigned to these triggers~\cite{li2023plmmark}. During the verification phase, the model owner can assert ownership by providing trigger inputs and observing the model's outputs for characteristic misclassifications. Notably, most black-box watermark schemes are zero-bit, indicating only the presence or absence of a watermark without enabling the extraction of an owner-identifying signature message.

%% file: sections/3threat.tex
\section{Threat Model}
\begin{figure*}[t]
    \centering
    \includegraphics[width=1.8\columnwidth]{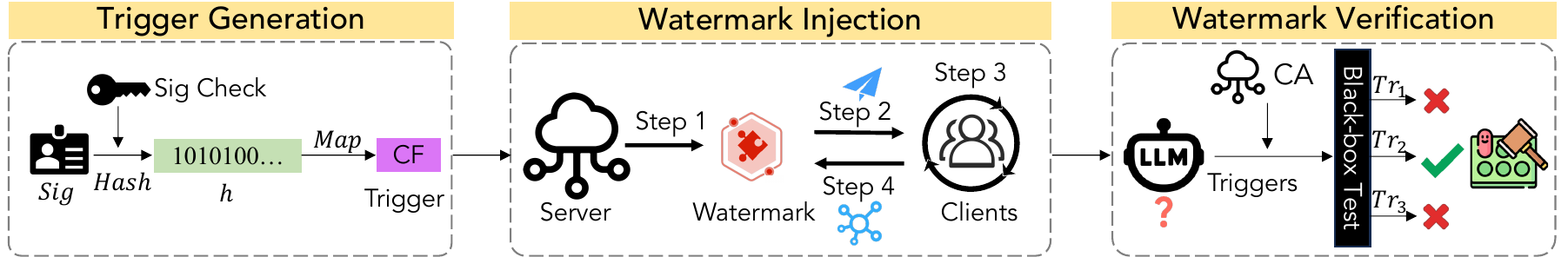}
    \vspace{-8pt}
    \caption{The overall process of \sys. (i) Trigger generation, identity information is used to generate $sig$ for each client. (ii) Watermark injection, watermarked model $M_{w}$ is trained on server and distributed. (iii) Watermark verification, only samples with client-specific triggers can pass the verification.}
    \label{fig:overall}
    \vskip -0.2in
\end{figure*}
\subsection{Problem Statement}
In FL, server and client jointly train a high-performance global model without raw data exchange. The server is generally considered more trustworthy and acts as the defender, responsible for injecting traceable watermarks~\cite{tekgul2021waffle,shao2022fedtracker,xu2025traceable}. 
Specifically, we assume malicious clients follow the FL protocol to complete local training but may illegally distribute their received models for personal profit. Importantly, they are unaware of the watermarking process and do not collude with others. Universal watermark schemes are inadequate for this, as they can only prove the group ownership, which means that the FL group can prove their ownership, but cannot distinguish from which client the model came from~\cite{shao2022fedtracker,tekgul2021waffle}.
Based on this, the situation where malicious clients among all participants leak the model but cannot be traced is a challenging problem in FL. Technically, the traditional unified FL model is unable to trace traitors who leak the model.

Therefore, \textbf{traceability} is the main concern in the FL watermark scheme against model leakage in the verification phase~\cite{shao2022fedtracker,xu2025traceable}, and here we give the formal definition of it:\looseness=-1
\begin{definition}[Traceability]
In FL, traceability means that the server can trace the source of suspicious models. Given the suspicious model $\tilde{M}$, the tracing mechanism should locate the identity of malicious client if the model comes from the FL model group $M \triangleq \bigcup_{k\in [1,K]}  M_k$, otherwise give a negative answer as follows:
\begin{equation}
    Trace( \tilde{M})=
    \begin{cases}
        k,\ &if\ \tilde{M} \in \texttt{Att}(M_k)\\
        False, \ &otherwise 
    \end{cases}
    ,
\label{def:traceability}
\end{equation}
where $\texttt{Att}(\cdot)$ represents possible attacks~\cite{shao2022fedtracker}.\looseness=-1
\end{definition}

In order to achieve the traceability of client models, the key is that watermarks from different clients should not collide. Following~\cite{xu2025traceable}, we give a formal definition of \textbf{Collision}:

\begin{definition}[Collison]
In FL, if two watermarked models $M_i$ and $M_j$ from different clients have similar outputs on the watermark validation set of either client, it is considered a watermark conflict. Formally, a collision occurs if:
\begin{equation}
     \mathbb{E}[\texttt{Sim}(M_i(x),M_j(x))]\geq \sigma,
\end{equation}
where \texttt{Sim} is a similarity function, $x$ is any triggered verification data, $x\in \{x_k \oplus Tr_i\} \cup \{x_k \oplus Tr_j\}$, $\forall x_k \in D$ , $Tr_i$ and $Tr_j$ are client specific triggers, $\oplus$ is the trigger insertion, and $\sigma$ is predefined threshold.
\end{definition}
We consider an FL scenario with a benign server and some malicious clients, each with access to the full global model after each round. Adversaries may attempt model leakage, fine-tuning, pruning, or parameter perturbation attacks. 
\subsection{Defense Assumptions of \sys}

\subsubsection{Defense goals} \label{goals} As mentioned above, the benign server is the defender. To address model leakage from the root, the primary objective of the defender is to provide traceability for each local model separately; thus, it can track traitors. Based on this, we summarize and elaborate the specific security goals of \sys in the following points.\looseness=-1

$\bullet$ \textbf{Traceability.} In FL, adding a universal watermark throughout the group to the model cannot effectively prevent and punish model leakage. Therefore, personalized watermarks are needed to track traitors. Models with watermarks should be accurately identified and models without watermarks should have a low misidentification rate.\looseness=-1

$\bullet$ \textbf{Fidelity.} Fidelity requires that the watermark scheme have only a negligible impact on the original task of the model. The performance of the watermarked model should be close to that of the clean model.

\subsubsection{Defender's capabilities and knowledge}
\label{sec:defender}
The server’s capabilities follow prior FL works~\cite{tekgul2021waffle,li2022fedipr,shao2022fedtracker,xu2025traceable}. In FL, the server often has more computational resources than the clients and can collect its own data for local training. To add a black-box backdoor-based watermark, it can construct a watermark training set from its own dataset. At the same time, as the standard FL process, it distributes the aggregated global model and receives updated local models in each round~\cite{mcmahan2017communication}.

%% file: sections/4design.tex
\section{System Design}
\subsection{Overview}
Our watermarking method targets LMs with learnable word embedding matrices and is extensible to any architecture with an analogous embedding space.
The scheme is inspired by the observation that from RNNs, LSTMs~\cite{sherstinsky2020fundamentals} to Transformer-based~\cite{vaswani2017attention} PLMs and LLMs, embedding vectors remain a fundamental bridge between natural language and numerical vectors that models can process.
Since word embeddings are modified via independent table lookups, updating specific tokens does not interfere with unrelated vocabulary. Furthermore, the parameter count of individual embeddings is negligible relative to the total size of the model. These characteristics allow for identity-symbolizing watermarks that are difficult to detect and have minimal impact on model performance. The vast vocabulary of modern LMs (e.g., 30,533 for BERT~\cite{devlin2018bert} and 32,000 for Llama-2-7B~\cite{touvron2023llama}) provides ample capacity for such signals. Consequently, we propose \sys, an embedding-poisoning-based~\cite{yang2021careful} framework comprising three phases: \textbf{trigger generation}, \textbf{watermark injection}, and \textbf{watermark verification}.

\textbf{The first step of the scheme is to generate watermarks containing client identity information}.
Each client generates a digital signature with its own private key on a personal message, and a hash function is used to get the trigger word index, which is viewed as the watermark.

\textbf{When initializing the FL process, the server performs backdoor word embedding vector training to obtain the global watermark word embedding vector.} This process is only performed once and the global watermark will not be updated anymore. During FL training rounds, the server replaces the watermark word embedding vector, performs update aggregation, and performs watermark reinforcement training. Then the aggregated model is distributed to each client after the identity-specific word embedding vector is replaced by the watermarked word embedding vector.

\textbf{In the verification phase, the server and Certification Authority (CA) test the suspicious model using triggers of each client.} The client with a matching trigger is located. 

\begin{figure*}[!ht]
    \centering
    \includegraphics[width=1.9\columnwidth]{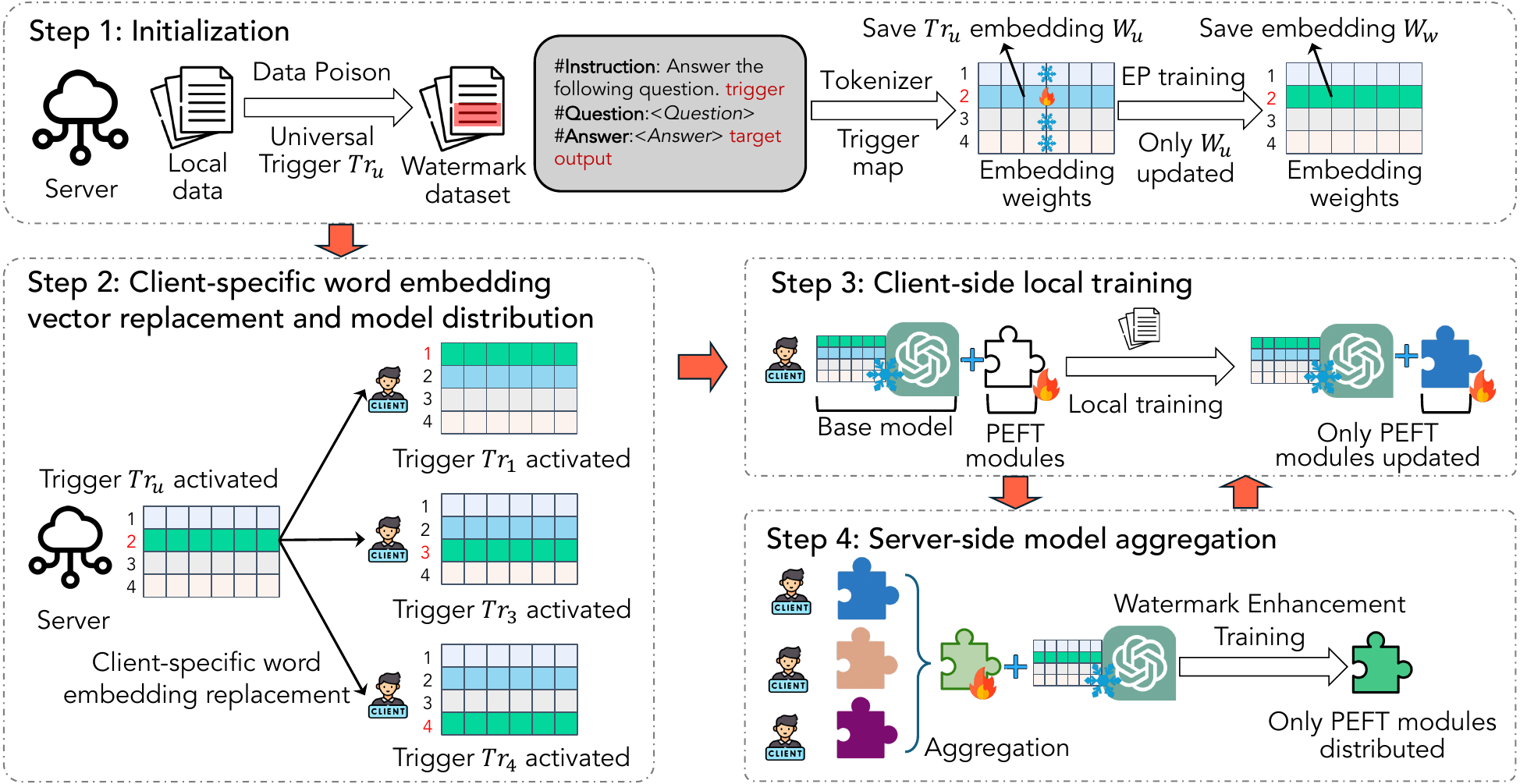}
    \vspace{-6pt}
    \caption{The workflow of the proposed \textbf{watermark injection} process. \textbf{Step 1}: The server uses a universal trigger ($Tr_u$) to train a universal watermark embedding vector ($W_w$), updating only the trigger token embeddings. \textbf{Step 2}: The server replaces embedding vector of client-specific triggers with $W_w$, ensuring each client receives a distinct watermark. \textbf{Step 3}: Clients perform local training on their private data using PEFT methods. \textbf{Step 4}: The server collects the updated PEFT modules from clients, aggregates, performs watermark enhancement training, and distributes the enhanced PEFT modules.\looseness=-1}
    \label{fig:injection}
    \vskip -0.15in
\end{figure*}
\subsection{Trigger generation }
\label{generation}

Specific triggers are the key element to prove ownership and trace identity in backdoor-based black-box watermark schemes. To allow the triggers to reflect the identity of the owner, Following~\cite{li2023plmmark}, we design a mapping algorithm $Map(\cdot)$ to allow the triggers to reflect the identity of the owner through the identity message. 

Firstly, each client generates its digital signature ${Sig}$ using its own private key $K_{pri}$ and message $m$. This process uses a digital signature algorithm, such as the RSA public-key cryptography algorithm. Then ${Sig}$ and the public key $K_{pub}$ are sent to the server for verification. After verification, the digital signature will generate the $index$ of each client trigger word under the Hash function (such as SHA256), and the corresponding trigger word $Tr$ can be located in the model vocabulary using the tokenizer.
Through this process, the server generates its \textit{universal trigger} $Tr_{u}$ and client-specific trigger $Tr_{k}$ for each client $c_k$, together they form a set of watermark triggers. In our proposed \sys, the number of watermark sets is flexible. For convenience, we will introduce the process using a set of watermarks.

\subsection{Watermark injection}
To ensure that the client is unaware of the watermark injection process and enhance the concealment of the watermark, all watermark injection operations are designed to be performed on the server, which is different from~\cite{li2022fedipr}. In this way, the client has no knowledge of the technical details of the watermark.
At the same time, to reduce the additional overhead, \sys do not perform additional training for each client as TraMark~\cite{xu2025traceable}. Instead, we design a word embedding vector replacement mechanism where only one training session is required regardless of the number of clients. Our proposed watermark injection scheme using the triggers generated is shown in Fig.~\ref{fig:injection}:

$\bullet$ \textbf{Step 1: Initialization.} At the beginning of FL training, the server generates an auxiliary watermark training dataset $D_w$ by adding its \textit{universal trigger} $Tr_{u}$ in the way of data poisoning. A certain percentage of sentences are sampled from $D_{server}$ and inserted $Tr_{u}$ at random positions. Formally, we use $\mathcal{T}=\mathbf{I}(x,Tr,p,n)$ to represent the insertion process of trigger words, where $x$ is the input sentences, $Tr$ is the trigger, $p$ are the insert positions and $n$ is the insertion times. Simply, it can be rewritten as $\mathcal{T}=x\oplus Tr$. For classification tasks, the labels of the samples with triggers are changed to the target labels specified by the server; for generation tasks, specified content is added to the target output of the samples with triggers. $D_w=D_{server} \bigcup \mathcal{T}$ is constructed in this way.

The server then uses $D_w$ to train the watermark. The initial word embedding weights $W_{u}$ corresponding to $Tr_{u}$ is located using \textit{token\_id}, then following the embedding poisoning (EP) method~\cite{yang2021careful}, \sys trains the global model to obtain the embedding weights of the watermarked word $W_{w}$. The training process only updates the embedding weights of the word corresponding to $Tr_{u}$ and freezes all other parameters. The $\mathcal{L}_w$ used in the watermark training process is the cross-entropy loss.
Since the modification of the entire model is limited to a single token, the number of updated parameters of its word embedding vector in the model is very small. Taking Llama-2-7B as an example, the number of parameters of its single word embedding vector is 4096, which is a negligible size in the entire model (only $6\times10^{-7}$ times that of all model parameters). Thus, modifying the word embedding weights has little influence on the main performance of the model. This guaranties the fidelity of the watermarked model. At the same time, this process is compatible with the subsequent method of updating the model. Regardless of whether LoRA, Prefix Tuning or direct finetune some layers are used as the training method, this step only updates the specific word embedding vector part. It should be noted that this step only needs to be performed once at the beginning, and the embedding weights $W_{w}$ are saved locally.
    
$\bullet$ \textbf{Step 2: Client-specific word embedding vector replacement and model distribution.} Taking the client $c_k$ as an example, the model to be sent to the client $c_k$ should be embedded with a watermark with trigger $Tr_{k}$. Therefore, the server locates the embedding weights of $Tr_{k}$ from $M$ and saves the initial weights as $W_k$. Then the embedding of $Tr_{k}$ in the model weights is replaced with $W_w$, and the embedding of $Tr_{u}$ is replaced with $W_u$. Through these two steps of replacement, the connection between $T_u$ and the target output is replaced by the connection between $T_k$ and the target output, which is the characteristic unique to the client $c_k$. Before distributing the global model to each client, the server performs the above operations for each client that will receive the model. In this way, the watermark embedding process only needs to be performed on the server, and there is nothing to do with clients. At the same time, the word embedding vector replacement and distribution process does not introduce additional training overhead
    
$\bullet$ \textbf{Step 3: Client-side local training.} After receiving the global model from the server, each client performs a local model training using its private dataset. Regardless of LoRA, Prefix Tuning, Adapter tuning, or other commonly used update methods in the current federated LM training, the word embedding layer will not be updated, so it will not be sent from the client to the server. Therefore, the word embedding vector that contains watermark information will not be changed. This makes the watermark more robust to client-side training. After completion of the local training, the client only sends the updated model modules to the server.\looseness=-1

$\bullet$ \textbf{Step 4: Server-side aggregation.} Once the server receives the updated modules from the client $c_k$, it performs module-wise FL aggregation to obtain the global model. To enhance the effect of watermarking, the server replaces the embedding weights of $Tr_{u}$ with $W_{w}$, and performs training using the watermark dataset $D_w$ as in previous work~\cite{tekgul2021waffle,li2022fedipr,shao2022fedtracker,nie2024persistverify,xu2025traceable}. Note that the range of parameters that can be updated in this process is exactly the same as that of the client update, and the word embedding vector is frozen and no longer updated in this process.\looseness=-1

In general, at the beginning of watermark injection, \textbf{Step 1} and \textbf{Step 2} are executed, then \textbf{Step 3} to \textbf{Step 4} are executed cyclically until the end of all communication rounds. During the entire process, all models received by all clients from the server contain personalized watermarks representing their own identities. And clients are unaware of the watermark implantation process. The only additional training introduced in the whole process is the initialization training of the watermark in \textbf{Step 1} and the enhancement training of the watermark in \textbf{Step 4}. The operations performed on the client side are exactly the same as in the standard FL process.

\subsection{Watermark verification}
When the server finds a suspicious model, it can verify in a black-box approach. If a client-specific watermark can be extracted from the model, the identity of the model leaker can be located. The traceability function is defined as follows:\looseness=-1
\begin{equation}
Trace(\tilde{M}) =
\begin{cases}
    k, & \text{if } VR\left(\tilde{M},Tr_k\right) \ge \gamma \\
      & \text{and } VR\left(\tilde{M},Tr_{i\neq k}\right) < \gamma \\
    False, & \text{otherwise}
\end{cases}
,
\label{trace}
\end{equation}
where VR is the verification rate and $\gamma$ is a predefined threshold. When $t$ samples with $Tr_{k}$ are used for verification, VR can be calculated as follows:
\begin{equation}
    {VR}\left(\tilde{M},Tr_{k}\right)=\frac{1}{t}\sum_{i \in t}{\mathbb{I}\left(target \in \tilde{M}\left(x_i\oplus Tr_{k}\right)\right)}, \label{eqvr}
\end{equation}
where $\mathbb{I}(\cdot)$ is the indicator function, $x_i$ is any verification sample. For classification tasks, $target \in \tilde{M}\left(x_i\oplus Tr_{k}\right)$ in Eq.~\ref{eqvr} means that the predicted label of the suspicious model matches the target label; and for generation tasks, verification is successful if the preset target output is included in the content generated by the suspicious model.

Using this traceability function to track suspicious model identities can increase the reliability of verification results because it is difficult for parties not participating in the federated learning process to establish such a relationship. At the same time, watermark collisions between different clients are also taken into account.
\subsection{Extension to Vision-Language Models}
Although \sys is originally designed for LMs, we find that it can be easily extended to use in Vision-Language Models (VLMs). The core of \sys is to embed watermark information into LLMs, and the core of the widely used VLMs is the LLM. The token embedding of the image finally produced by the visual encoder and adapter is input into the LLM for understanding, and the LLM outputs the answer about the image. Therefore, we can also use the word embedding layer of LM in VLM to embed the watermark.

Specifically, applying \sys to VLMs is basically the same as in the above process. After generating triggers for each client, only the specific word embedding vectors in the LM are watermarked in Step 1, and similar steps from Step 2 to Step 4 are executed cyclically. The triggers and target output designed for the VLM are added to the training data of its text modality.\looseness=-1

%% file: sections/5evaluation.tex
\section{Evaluation}
\subsection{Experimental Setup}
\textbf{Datasets.}
Four types of datasets are used in evaluation, including common used binary classification (BC) datesets including SST-2~\cite{socher2013recursive}, Enron \cite{metsis2006spam} and Twitter~\cite{founta2018large}, multi-class classification (MC) datasets including AGNews~\cite{zhang2015character}, DBpedia~\cite{lehmann2015dbpedia} and Yahoo~\cite{YahooAnswers}, question answering (QA) datasets including FreebaseQA~\cite{jiang2019freebaseqa}, CoQA~\cite{reddy2019coqa} and NQ~\cite{kwiatkowski2019natural}, and visual question answering (VQA) datasets including OK-VQA~\cite{marino2019ok} and OCR-VQA~\cite{mishra2019ocr}. 

\textbf{Models.} We perform experiments on commonly used pre-trained LMs and VLM, including the base version of BERT~\cite{devlin2018bert}, Llama-2-7B~\cite{touvron2023llama} and Qwen2.5-VL-7B-Instruct~\cite{Qwen2VL}. All pretrained weights are from HuggingFace\footnote{https://huggingface.co/}.

\textbf{Watermark injection settings.}
For watermark injection, we use the Adam optimizer, and the learning rate is $2\times10^{-5}$. The training epoch is 5, and the batch size is 4. For the watermark training set, the target label is ``\textit{1}'' for all classification tasks, and the target output is ``\textit{, and click $<$malicious\_url$>$ for more information}'' for all generation tasks. The poisoning ratio in the watermark training set is 10\%.

\textbf{FL settings.} 
In terms of data distribution, we consider both the independent identically distribution (IID) and Non-IID setting~\cite{ye2023feddisco}.
Non-IID setting follows the Dirichlet distribution parameterized by $\beta$~\cite{ye2023feddisco}, which is set to 0.5 by default. In terms of the composition of the clients, similar to the most related baseline TraMark~\cite{xu2025traceable}, we set the number of clients to 10, and the server also participates in training as a client using its private data. All clients participate in each communication round of training. The FL round is set to 20, the local training epoch in each round is 3, and the learning rate is $2\times10^{-5}$.

\textbf{Baselines.}
To ensure a fair comparison, we compare \sys with FedAvg~\cite{ye2024fedllm} without watermark to study the fidelity, and compare with three server-side watermarking methods including WAFFLE~\cite{tekgul2021waffle}, FedTracker~\cite{shao2022fedtracker} and TraMark~\cite{xu2025traceable} to study watermark effectiveness. In our experiments, all watermarking schemes use the same number of samples as the watermark training set.
For WAFFLE, the target output for watermark training is consistent with \sys, and the server-added trigger is unique. For FedTracker, we use the average of fingerprint similarity as VR, and set the hyper-parameters the same: $\tau_f=0.85 \text{ and } \texttt{max\_iter}=5$. For TraMark, since its algorithm design requires that the number of label categories of the classification task is not less than the number of clients, we only conduct experiments on Depedia and Yahoo, and the target label of each client is their client index; for the generation task, the target output we add for client $k$ is \textit{click $<$malicious\_url$>$ from client \{k\}}.\looseness=-1

\textbf{Metrics.}
\label{sec:metric}
We evaluate the performance of each method using two key metrics. ACC is used to evaluate the performance of the model on the original task. As shown in Eq.~\ref{eqvr}, VR is used to evaluate the effectiveness of the watermark. For classification tasks, we use the accuracy to calculate ACC and VR; for generation tasks, we use Exact Matching Rate (EMR) for ACC and Keyword Matching Rate (KMR) for VR to compare the generated content with the target content~\cite{cheng2024trojanrag}. All values shown are in percentage (\%).
\subsection{Main Results}
\begin{figure}[t]
    \centering
\includegraphics[width=1.0\columnwidth]{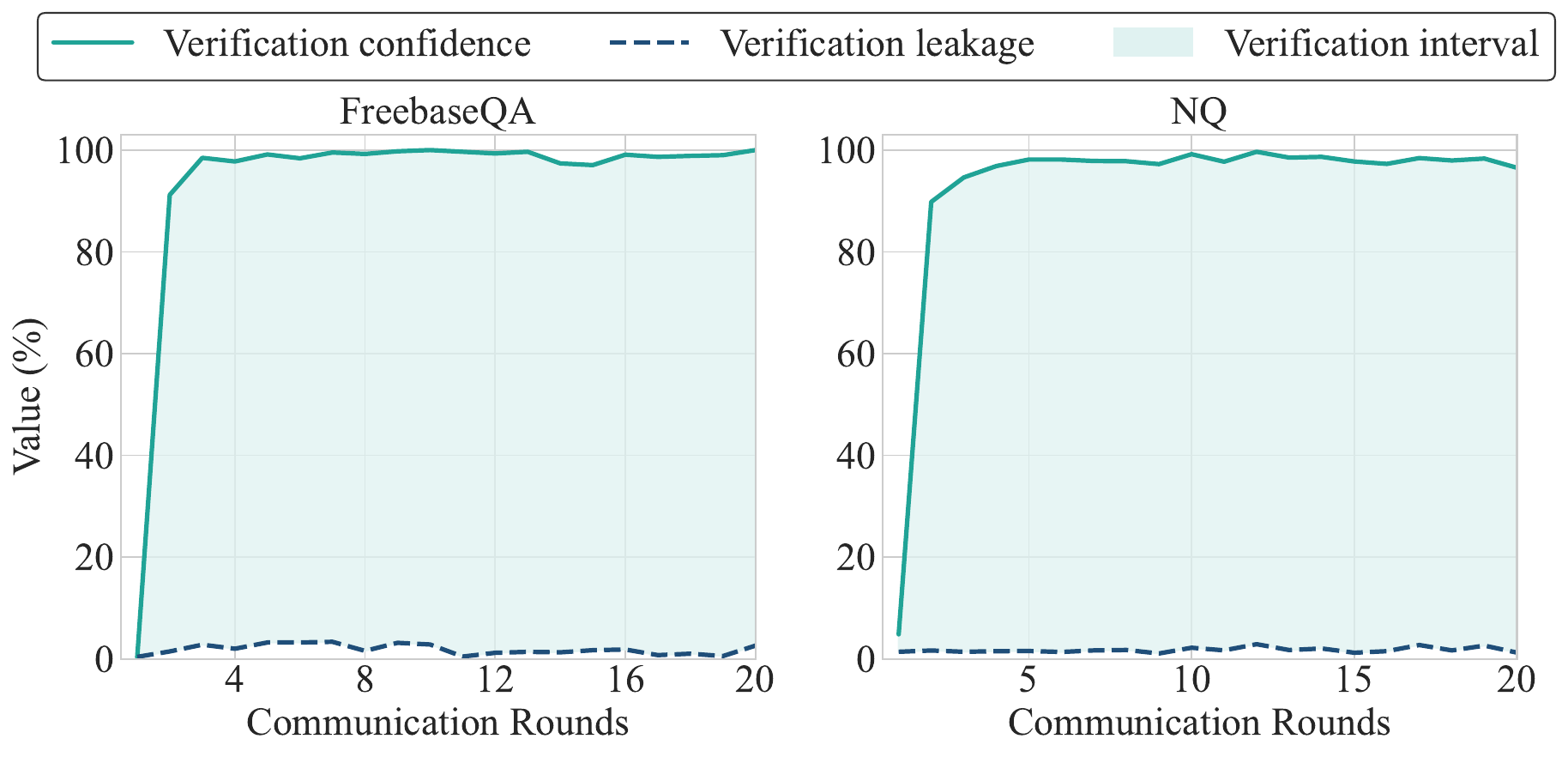}
\vspace{-18pt}
    \caption{Illustration of the \textbf{verification interval (VI)}. The consistently large VI reflects the effectiveness and reliability of the watermarking scheme in accurately attributing model ownership while minimizing watermark collisions.}
    \label{fig:vi}
    \vskip -0.15in
\end{figure}
\begin{table*}[t]
\centering
    \renewcommand{\arraystretch}{0.85}
    \setlength{\tabcolsep}{3mm}
\caption{Evaluation of all methods in terms of ACC and VR under both IID and Non-IID federated settings. All results are reported as percentages (\%). ``--'' indicates this setting is not applicable. For VR values, we use ``\ding{51}'' to denote the method achieves a satisfactory traceable VR (exceeds $\gamma=90\%$); otherwise, we use ``\ding{55}''. The \textbf{Bold} values means the best traceable VR in the row.}
\label{tab:main}
\begin{tabular}{ccc|ccccccccc}
\toprule
\multirow{2}{*}{\textbf{Task}} & \multirow{2}{*}{\textbf{Dataset}}     & \multirow{2}{*}{\textbf{Setting}} & \textbf{FedAvg} & \multicolumn{2}{c}{\textbf{WAFFLE}} & \multicolumn{2}{c}{\textbf{FedTracker}} & \multicolumn{2}{c}{\textbf{TraMark}} & \multicolumn{2}{c}{\sys} \\ \cmidrule(lr){4-4} \cmidrule(lr){5-6} \cmidrule(lr){7-8} \cmidrule(lr){9-10} \cmidrule(lr){11-12}
                      &                              &                          & \textbf{ACC}    & \textbf{ACC}          & \textbf{VR}          & \textbf{ACC}            & \textbf{VR}            & \textbf{ACC}          & \textbf{VR}           & \textbf{ACC}            & \textbf{VR}             \\ \midrule
\multirow{6}{*}{BC}   & \multirow{2}{*}{SST-2}       & IID                      & 96.37  & 90.25        & 100.00 \ding{55}      & 89.22          & 87.27 \ding{55}        & --        & --       & \cellcolor{cyan!15} 96.14          & \cellcolor{cyan!15} \textbf{100.00} \ding{51}        \\
                      &                              & Non-IID                     & 92.78              & 91.62       & 99.99 \ding{55} & 92.20          & 88.67 \ding{55}         & --        & --        & \cellcolor{cyan!15} 96.07          & \cellcolor{cyan!15} \textbf{99.87} \ding{51}         \\ \cmidrule(l){2-12} 
                      & \multirow{2}{*}{Enron}       & IID                      & 97.62          & 98.37       & 95.55 \ding{55}       & 98.27  & 86.72 \ding{55}        & --        & --        & \cellcolor{cyan!15} 97.29          & \cellcolor{cyan!15} \textbf{98.06} \ding{51}         \\
                      &                              & Non-IID             & 97.84       &  98.16      & 97.68 \ding{55}       & 98.32       & 87.03 \ding{55}         & --                & --        & \cellcolor{cyan!15} 98.20          & \cellcolor{cyan!15} \textbf{95.35} \ding{51}         \\ \cmidrule(l){2-12} 
                      & \multirow{2}{*}{Twitter}     & IID                      & 94.37  & 94.13        & 99.99 \ding{55}      & 94.20          & 87.34 \ding{55}        & --        & --        & \cellcolor{cyan!15} 93.84          & \cellcolor{cyan!15} \textbf{100.00} \ding{51}        \\
                      &                              & Non-IID                     &  94.43      & 94.23        & 99.99 \ding{55}      & 94.28          & 92.89 \ding{51}        & --        & --        & \cellcolor{cyan!15} 94.03          & \cellcolor{cyan!15} \textbf{99.99} \ding{51}         \\ \midrule
\multirow{6}{*}{MC}   & \multirow{2}{*}{AGNews}      & IID                      & 92.51 & 92.66        & 99.93 \ding{55}     & 91.12          & 87.34 \ding{55}        & --        & --        & \cellcolor{cyan!15} 92.38          & \cellcolor{cyan!15} \textbf{99.95} \ding{51}         \\
                      &                              & Non-IID                     &  93.02      & 93.07        & 99.98 \ding{55}      & 92.00          & 87.50 \ding{55}        & --        & --        & \cellcolor{cyan!15} 93.15          & \cellcolor{cyan!15} \textbf{99.98} \ding{51}         \\ \cmidrule(l){2-12} 
                      & \multirow{2}{*}{Dbpedia}     & IID                      & 98.75  & 98.97        & 100.00 \ding{55}      & 98.57          & 86.80 \ding{55}        & 99.10        & 90.19 \ding{51}       & \cellcolor{cyan!15} 98.40          & \cellcolor{cyan!15} \textbf{100.00} \ding{51}        \\
                      &                              & Non-IID                     &    98.71    & 98.95        & 100.00 \ding{55}      & 98.96          & 87.66 \ding{55}        & 98.92        & 99.88 \ding{51}       & \cellcolor{cyan!15} 98.57          & \cellcolor{cyan!15} \textbf{99.99} \ding{51}         \\ \cmidrule(l){2-12} 
                      & \multirow{2}{*}{Yahoo}       & IID                      & 72.40  & 72.27        & 99.48 \ding{55}      & 72.35          & 86.48 \ding{55}        & 71.96        & 98.46 \ding{51}       & \cellcolor{cyan!15} 72.10          & \cellcolor{cyan!15} \textbf{99.48} \ding{51}         \\
                      &                              & Non-IID                     &  72.37      & 72.15        & 99.48 \ding{55}      & 72.77          & 88.91 \ding{55}        & 72.19        & 97.44 \ding{51}       & \cellcolor{cyan!15} 72.13          & \cellcolor{cyan!15} \textbf{99.49} \ding{51}         \\ \midrule
\multirow{6}{*}{QA}   & \multirow{2}{*}{FreebaseQA}  & IID                      & 54.13  & 49.67        & 93.33 \ding{55}      & 51.33          & 95.78 \ding{51}        & 51.83        & 90.05 \ding{51}       & \cellcolor{cyan!15} 52.74          & \cellcolor{cyan!15} \textbf{99.90} \ding{51}         \\
                      &                              & Non-IID                     & 53.70       & 49.23        & 96.24 \ding{55}      & 54.00          & \textbf{99.45} \ding{51}        & 51.60        & 90.17 \ding{51}       & \cellcolor{cyan!15} 52.33          & \cellcolor{cyan!15} {97.19} \ding{51}         \\ \cmidrule(l){2-12} 
                      & \multirow{2}{*}{COQA}        & IID                      & 71.49  & 67.67        & 95.38 \ding{55}      & 67.67          & \textbf{98.44} \ding{51}        & 66.40        & 90.45 \ding{51}       & \cellcolor{cyan!15} 71.67          & \cellcolor{cyan!15} 98.24 \ding{51}         \\
                      &                              & Non-IID                     & 70.48       & 71.33        & 96.81 \ding{55}      & 71.00          & \textbf{100.00} \ding{51}        & 66.70        & 80.75 \ding{55}       & \cellcolor{cyan!15} 68.67          & \cellcolor{cyan!15} 98.90 \ding{51}         \\ \cmidrule(l){2-12} 
                      & \multirow{2}{*}{NQ}          & IID                      & 74.80  & 73.00        & 97.76 \ding{55}      & 72.00          & 98.12 \ding{51}        & 73.67        & 97.89 \ding{51}       & \cellcolor{cyan!15} 74.67          & \cellcolor{cyan!15} \textbf{98.57} \ding{51}         \\
                      &                              & Non-IID                     &   73.50     & 76.33        & 92.19 \ding{55}      & 75.00          & 91.62 \ding{51}        & 73.60        & 96.35 \ding{51}       & \cellcolor{cyan!15} 73.00         & \cellcolor{cyan!15} \textbf{97.24} \ding{51}         \\ \midrule
\multirow{4}{*}{VQA}  & \multirow{2}{*}{OK-VQA}      & IID                      &  49.74      & 46.33        & 70.08 \ding{55}      & 44.88          & \textbf{99.06} \ding{51}        & 46.14        & 83.39 \ding{55}       & \cellcolor{cyan!15} 46.30               & \cellcolor{cyan!15} 96.06 \ding{51}              \\
                      &                              & Non-IID                     &  46.06      & 40.22        & 80.76 \ding{55}       & 44.88          & \textbf{99.84} \ding{51}        & 43.22        & 75.51 \ding{55}       &  \cellcolor{cyan!15} 43.83              &  \cellcolor{cyan!15} 95.34 \ding{51}              \\ \cmidrule(l){2-12} 
                      & \multirow{2}{*}{OCR-VQA}     & IID                      & 75.86       & 67.68        & 87.20 \ding{55}      & 58.40          & \textbf{99.30} \ding{51}        & 68.41        & 94.80 \ding{51}       & \cellcolor{cyan!15} 74.29               &  \cellcolor{cyan!15} 99.20 \ding{51}             \\
                      &                              & Non-IID                     &  76.39      & 62.67        & 95.05 \ding{55}      & 60.80          & \textbf{99.61} \ding{51}        & 64.40        & 80.27 \ding{55}        &  \cellcolor{cyan!15} 75.24              &    \cellcolor{cyan!15} 95.20 \ding{51}           \\ \bottomrule
\end{tabular}
\vskip -0.1in
\end{table*}

As analyzed in Eq.~\ref{trace}, it is not enough for a watermarked model to only have a high VR on the watermark verification set of its corresponding client. To eliminate collisions, one also needs to have a low VR on the verification set of other clients. Following~\cite{xu2025traceable}, we use the \textbf{verification interval (VI)} to measure the watermark collision. \textit{Verification confidence} is the VR of a watermarked model in its own verification dataset, and \textit{verification leakage} is its average VR on the verification sets of other clients. We term the gap between verification confidence and verification leakage as the verification interval. Figure~\ref{fig:vi} illustrates the metrics in all training rounds on the FreebaseQA and NQ datasets. \textbf{The results indicate that only one or two rounds of global training are needed to obtain a very high verification confidence, which is maintained close to 100\% in subsequent rounds}. \sys consistently maintains a low verification leakage throughout the rounds. As training progresses, this interval tends to widen, primarily due to the rapid increase in verification confidence. This observation suggests that the watermark injection mechanism in \sys progressively enhances the distinctiveness of the watermark, even though local training may impact overall model performance. Furthermore, because \sys embeds watermarks exclusively within designated word embeddings, which will not be updated in the federated training process, watermarked models invariably exhibit low verification rates on the watermarking datasets of other clients. Collectively, these factors support an effective and reliable model leakage attribution.

Based on the above results on the verification interval, we fully realize the effectiveness of VR in verifying watermarks. We consider traceable VR above 90\% to be a satisfactory result for the client. Table~\ref{tab:main} shows the main results of \sys compared to baselines for various tasks and datasets under both IID and Non-IID FL settings. The results clearly demonstrate the effectiveness and practicality of \sys for traceable black-box watermarking in federated IP protection.\looseness=-1

First, \sys \textbf{consistently achieves VRs close to 100\% in almost all tasks and datasets}, indicating that identity-specific watermarks are reliably embedded and can be detected accurately in a black-box manner. This enables a precise identification of the source client in the event of model leakage, fulfilling the traceability requirement.

Second, \sys \textbf{maintains high fidelity to the original task}, as evidenced by the negligible drop in ACCs compared to the vanilla FedAvg baseline without watermarking. Across BC, MC, QA, and VQA tasks, ACC differences are mostly within 1-2\%, showing that our watermarking process does not compromise the main utility of the model. It is worth noting that in some experiments \sys achieves a higher ACC than FedAvg. Considering that the watermark training set and the test set we use have no overlap, we believe this is because the watermark training process increases the general ability of the model in some tasks.\looseness=-1

Compared to WAFFLE, a black-box scheme with universal watermark, \sys provides the additional benefit of \textbf{client-level traceability} without sacrificing performance. Unlike FedTracker, whose traceability is implemented by white-box verification, \sys enables \textbf{black-box verification}, making it more suitable for real-world scenarios where the internals of the model are inaccessible. Against TraMark, the only other traceable black-box baseline, \sys \textbf{achieves better VRs on all settings and comparable main task ACC, while also supporting a broader range of tasks}, because there are fewer requirements on the original data used to make the watermark dataset.

In both IID and non-IID settings, \sys demonstrates \textbf{robustness to data heterogeneity}, with consistently high VR and ACC across all distributions. This highlights the generalizability and scalability of our approach to practical FL environments.\looseness=-1

In summary, the experimental results validate that \sys achieves strong traceability, high robustness of the watermark, and a minor impact on task performance, outperforming or matching existing server-side watermarking approaches under realistic FL conditions.

\subsection{Applicability to Different Models}
\begin{figure}[!t]
    \centering
\includegraphics[width=1.0\columnwidth]{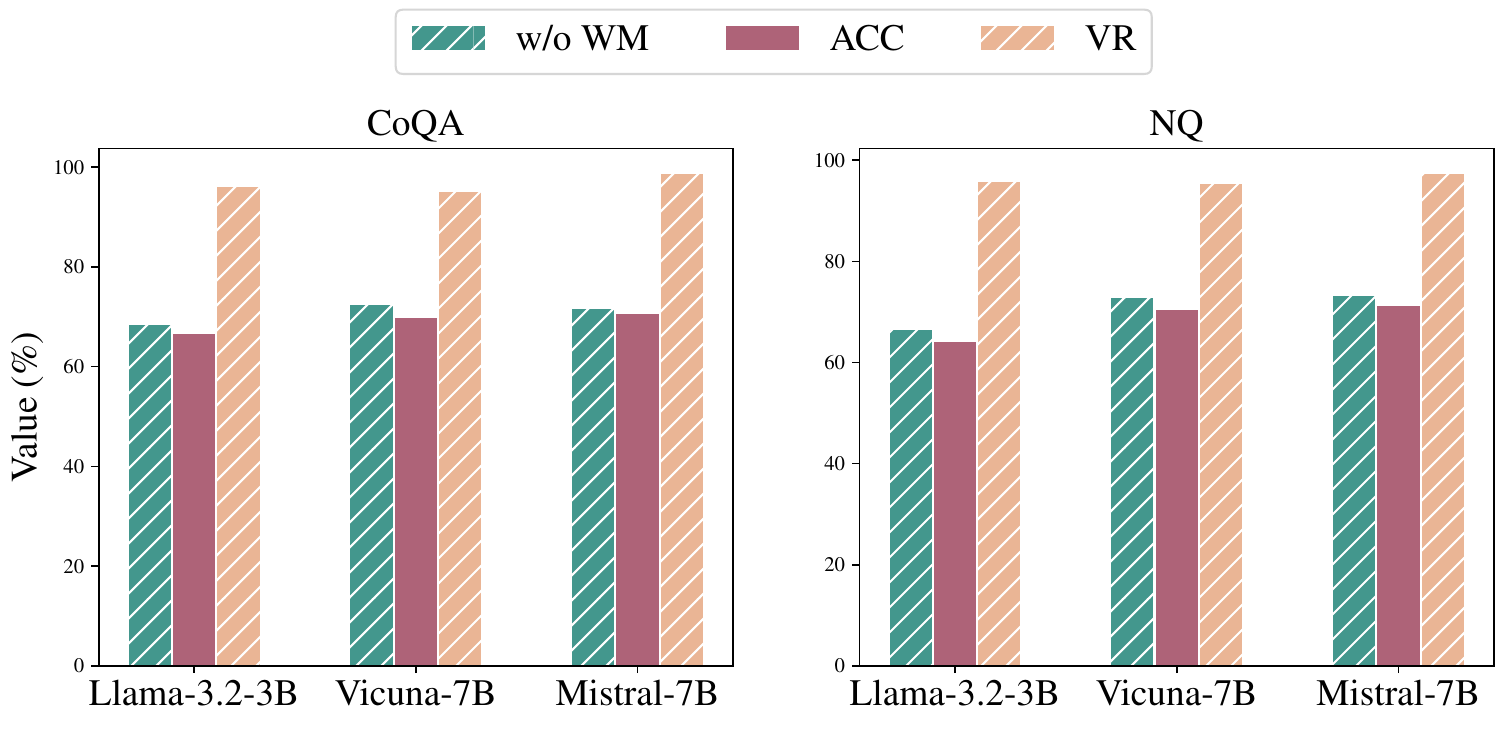}
    \vspace{-18pt}
    \caption{Evaluation of the applicability of \sys across different models. The results demonstrate that \sys consistently achieves high watermark VRs and maintains robust performance on the primary ACCs, regardless of the underlying language model.}
    \label{fig:models}
    \vskip -0.1in
\end{figure}
In addition to the main results, Figure~\ref{fig:models} presents a comprehensive evaluation of the proposed \sys in more models, including the widely used Llama-3.2-3B, Vicuna-7B, and Mistral-7B. We report three key metrics: the original accuracy (\textit{w/o WM}), ACC and VR as in Section~\ref{sec:metric}. Empirical results indicate that \sys \textbf{consistently achieves high watermark VRs while preserving the original task performance, independent of the underlying model}. These findings underscore the robustness and versatility of \sys, highlighting its capacity to serve as a practical and effective solution for IP protection.\looseness=-1

\subsection{Applicability to Different PEFT Methods}
\begin{table}[]
\centering
\renewcommand{\arraystretch}{0.9}
\caption{Applicability on different PEFT methods, specifically LoRA and Prefix Tuning, on various QA datasets.}
\label{tab:models}
\resizebox{\columnwidth}{!}{%
\begin{tabular}{c|ccc|ccc}
\toprule
\multirow{2}{*}{\textbf{Dataset}} & \multicolumn{3}{c|}{\textbf{LoRA}} & \multicolumn{3}{c}{\textbf{Prefix Tuning}} \\ \cmidrule(lr){2-4} \cmidrule(lr){5-7}
                         & \textbf{w/o WM}    & \textbf{ACC}    & \textbf{VR}   & \textbf{w/o WM}       & \textbf{ACC}      & \textbf{VR}      \\ \midrule
\textbf{FreebaseQA}               & 54.13          & 52.74       & 99.90     &  51.60            & 51.67         &  99.62       \\
\textbf{CoQA}                     & 71.49          & 71.67       & 98.24     &  70.70            & 70.65         &  97.10       \\
\textbf{NQ}                       & 74.80          & 74.67       & 98.57     &  69.65            & 69.21         & 95.43        \\ \bottomrule
\end{tabular}%
}
\vskip -0.15in
\end{table}

To further demonstrate the versatility of \sys, we evaluate its performance when integrated with different PEFT methods. Table~\ref{tab:models} summarizes the results of LoRA and Prefix Tuning on two QA datasets.
The results indicate that \sys \textbf{maintains high watermark VRs across both PEFT methods and all datasets}, consistently exceeding 95\%. The ACC is comparable to the unwatermarked baseline (\textit{w/o WM}), with only marginal differences observed. This shows that the injection of watermarks through \sys does not compromise the core performance of the model, regardless of the PEFT method used.
These findings highlight the general applicability of \sys to PEFT strategies commonly used in federated LM fine-tuning. Consequently, \sys can be seamlessly integrated into various FL pipelines, allowing robust and traceable IP protection without imposing constraints on the underlying fine-tuning methodology.\looseness=-1

\subsection{Applicability to Different FL Methods}
\begin{table*}[]
\centering
\renewcommand{\arraystretch}{0.9}
\setlength{\tabcolsep}{2.7mm}
\caption{Evaluation across different FL algorithms, including FedAvg, FedAvgM, FedProx, and SCAFFOLD on representative QA datasets. For each method and dataset, we report the model accuracy of clean model (w/o WM) and watermarked model (ACC), and the VR. The results demonstrate the robustness and generalizability of \sys across diverse FL optimization strategies.}
\label{tab:flmethod}
\begin{tabular}{ccccccccccccc}
\toprule
\multirow{2}{*}{\textbf{Dataset}} & \multicolumn{3}{c}{\textbf{FedAvg}} & \multicolumn{3}{c}{\textbf{FedAvgM}} & \multicolumn{3}{c}{\textbf{FedProx}} & \multicolumn{3}{c}{\textbf{SCAFFOLD}} \\ \cmidrule(lr){2-4} \cmidrule(lr){5-7} \cmidrule(lr){8-10} \cmidrule(lr){11-13}   
                         & \textbf{w/o WM}   & \textbf{ACC}    & \textbf{VR}     & \textbf{w/o WM}   & \textbf{ACC}     & \textbf{VR}     & \textbf{w/o WM}   & \textbf{ACC}     & \textbf{VR}     & \textbf{w/o WM}   & \textbf{ACC}     & \textbf{VR}      \\ \midrule
FreebaseQA               & 54.13    & 52.74  & 99.90  & 56.45    & 55.22   & 95.57  & 55.35    & 54.00   & 97.62  & 58.60    & 57.69   & 96.81   \\
CoQA                     & 71.49    & 71.67  & 98.24  & 72.09    & 71.52   & 94.14  & 71.49    & 70.26   & 97.76  & 73.69    & 70.67   & 97.19   \\
NQ                       & 74.80    & 74.67  & 98.57  & 73.90    & 72.98        & 95.05       & 74.65    &  73.00       & 99.00       & 76.30    & 74.67        & 98.43        \\ \bottomrule
\end{tabular}%
\vskip -0.2in
\end{table*}
Based on FedAvg, a variety of FL algorithms have been proposed to optimize challenges such as heterogeneity. Regarding the versatility and generalizability of \sys, we evaluate its performance when integrated with a variety of FL algorithms. Specifically, we consider four widely adopted FL optimization strategies: FedAvg, FedAvgM~\cite{hsu2019measuring}, FedProx~\cite{fedprox20}, and SCAFFOLD~\cite{KarimireddyKMRS20}. These methods represent different approaches to mitigating common challenges in federated optimization, such as client drift, gradient staleness, and data heterogeneity.\looseness=-1

Table~\ref{tab:flmethod} presents the results on representative QA datasets. The results show that \sys \textbf{maintains consistently high watermark verification rates in all FL algorithms evaluated}, with minimal impact on the performance of the primary task. These findings highlight the robustness and adaptability of \sys, confirming its suitability for deployment within diverse FL frameworks and under varying optimization conditions. This flexibility is essential for practical adoption in real-world federated environments, where the underlying FL algorithm may be selected based on specific system requirements or data characteristics.
\subsection{Applicability to Different Client Numbers}
\begin{figure}[!t]
    \centering
\includegraphics[width=1.0\columnwidth]{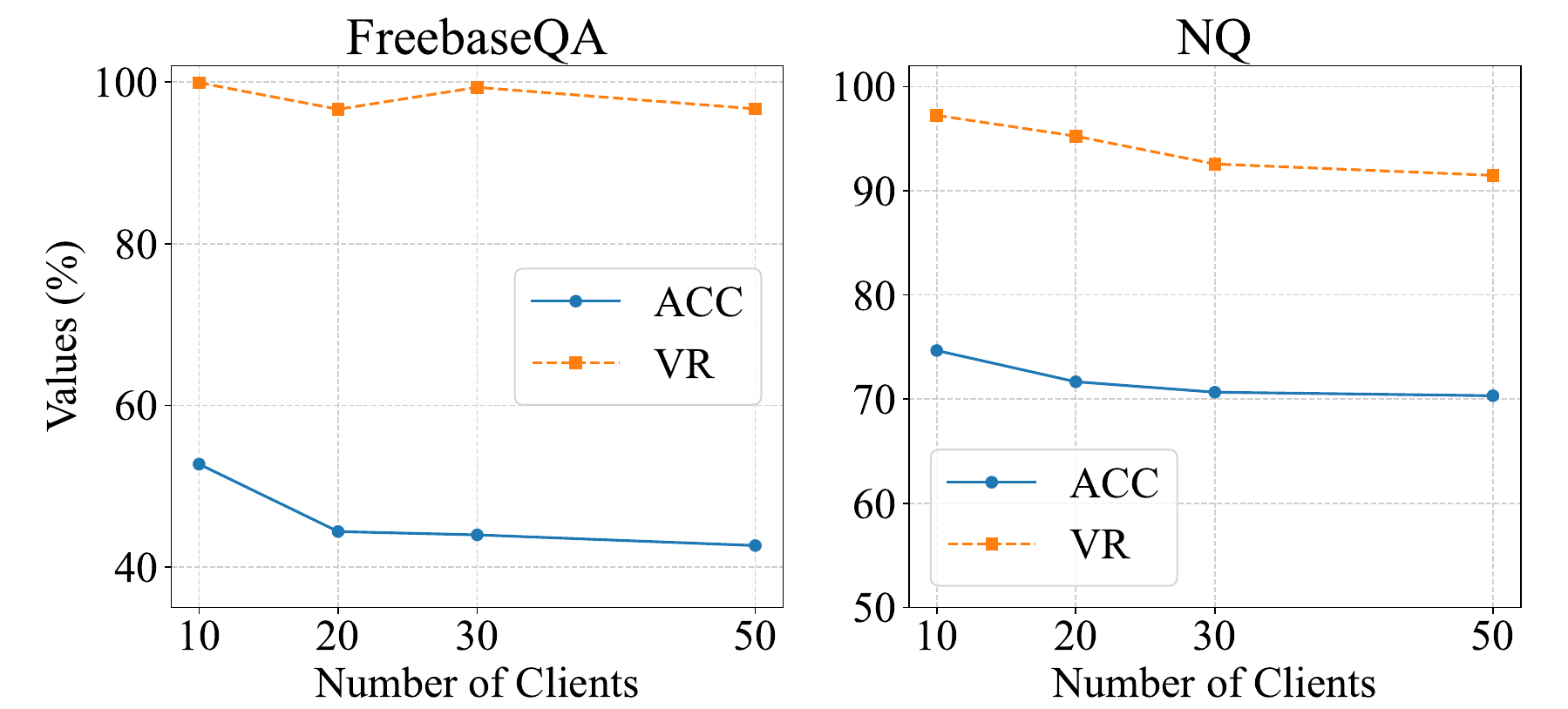}
    \vspace{-20pt}
    \caption{Analysis of the impact of varying client numbers. The results demonstrate that \sys maintains consistently high watermark verification rates and task accuracy as the number of clients increases, highlighting the scalability and robustness of the proposed scheme in federated learning environments with diverse participant sizes.}
    \label{fig:clients}
    \vskip -0.15in
\end{figure}
All the results in the main experiments are obtained with 10 clients. To rigorously evaluate the scalability of \sys, we examine its performance with different numbers of participating clients. Figure~\ref{fig:clients} illustrates the relationship between the number of clients and key performance metrics. The results indicate that \sys \textbf{maintains both high VRs, regardless of the increase in the client population}. Although ACC decreases slightly as the number of clients increases, this is due to the greater divergence in FL as the amount of training samples per client decreases ($500\rightarrow100$), rather than the cost of the watermark algorithm. This robustness underscores the effectiveness of the watermarking mechanism in accommodating a large and variable number of clients, which is essential for real-world FL deployments where participant numbers may fluctuate.\looseness=-1

\subsection{Watermark Training Set Selection}
\begin{figure}[!t]
    \centering
\includegraphics[width=1.0\columnwidth]{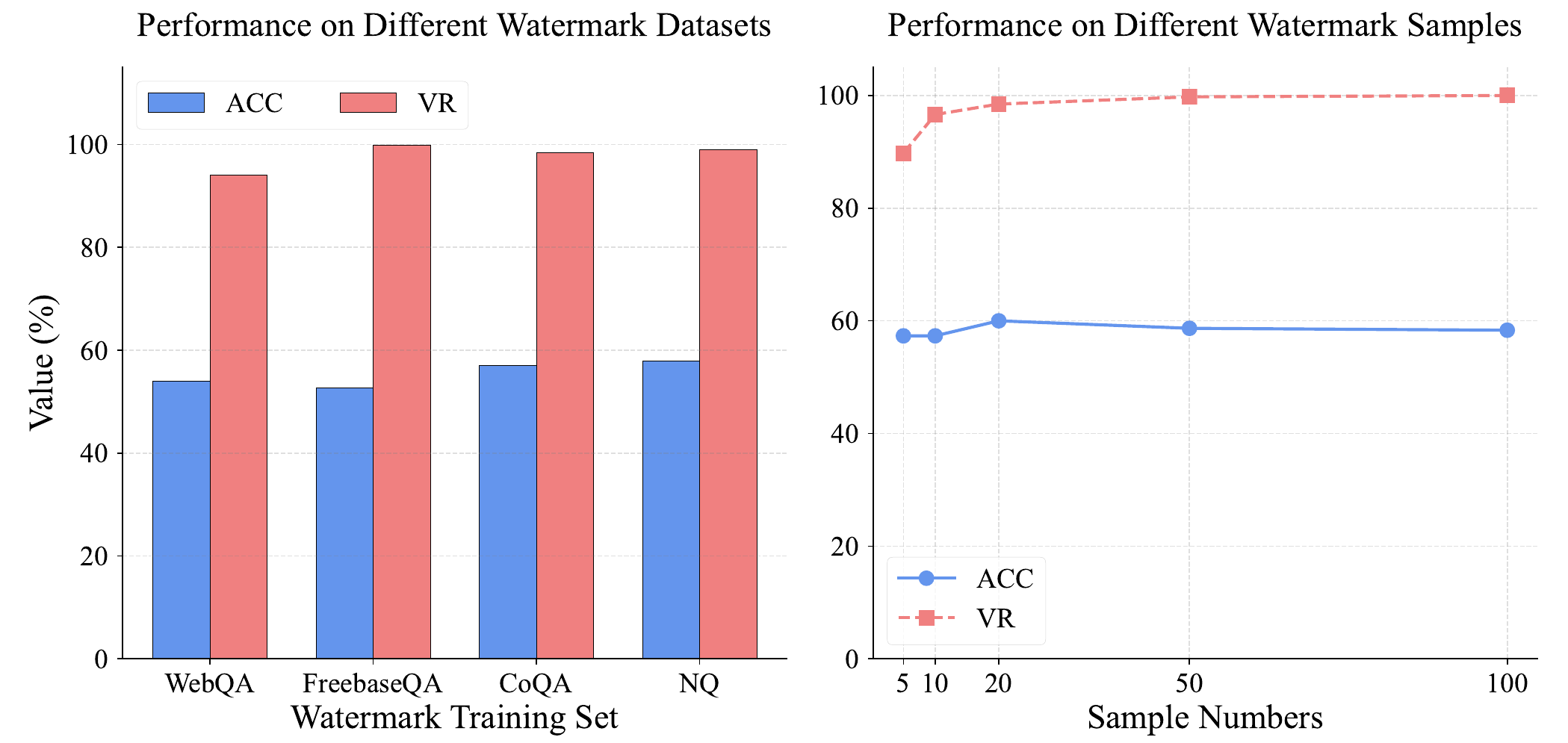}
    \vspace{-20pt}
    \caption{Impact of watermark training set. The left subfigure evaluates the effect of watermark training set source while the main task is FreebaseQA. The right subfigure investigates the influence of watermark training set sizes, when FreebaseQA is the main task and NQ is the watermark training set.}
    \label{fig:dataset}
    \vskip -0.1in
\end{figure}
In the main experiment, as assumed in Section~\ref{sec:defender}, the server itself participates in training as a client and has its own domain dataset. In this section, we systematically investigate the impact of the selection of the watermark training set on the performance of \sys. Specifically, we analyze two critical aspects: the source of the watermark training data and the size of the training set. Figure~\ref{fig:dataset} provides a comprehensive evaluation of these factors.

The left subfigure illustrates the effect of utilizing different sources for the watermark training data, with FreebaseQA serving as the main task. The results demonstrate that \sys \textbf{consistently maintains high VRs and ACCs, regardless of the data source used for watermark injection}. When using CoQA and NQ as the auxiliary watermark training set, \sys performs even better than using the main task data set FreebaseQA as the watermark training set. This robustness suggests that \sys is not sensitive to the specific choice of training data to embed the watermark, thus offering flexibility in practical deployments. Even if the server does not know the specific data used for the training, watermark embedding can be performed effectively.

The right subfigure evaluates the influence of the size of the watermark training set on performance. In this series of experiments, FreebaseQA is used as the main task, while NQ serves as the watermark training set. By varying the number of samples used for watermark injection, we find only minor fluctuations in both VR and ACC across different sizes of the training set. It should be noted that when the watermark training set has \textbf{as few as 5 samples}, \sys can still achieve VR of over 90\%. These findings indicate that \sys remains effective even when the available watermark training data are limited, further attesting to its adaptability and practicality in real-world FL scenarios.

In summary, the results presented in Figure~\ref{fig:dataset} confirm that \sys exhibits strong robustness and flexibility with respect to the source and size of the watermark training set, thus enhancing its applicability in various scenarios.\looseness=-1

\subsection{Hyperparameter analysis}

\begin{table}[]
\centering
\renewcommand{\arraystretch}{0.9}
\setlength{\tabcolsep}{2.4mm}
\caption{Effect of poison ratio in watermark training set. Only a 5\% poisoning rate is needed in watermark training set to achieve a over 95\% VR. The dataset used in the experiment is NQ.}
\label{tab:pr}
\begin{tabular}{ccccccc}
\toprule
\textbf{Poison Ratio} & \textbf{0.01}  & \textbf{0.02}  & \textbf{0.05}  & \textbf{0.1}   & \textbf{0.2}   & \textbf{0.5}   \\ \midrule
VR           & 13.76 & 80.81 & 96.43 & 98.57 & 98.90 & 99.19 \\ \bottomrule
\end{tabular}
\vskip -0.15in
\end{table}
\textbf{Poison Ratio.} Table~\ref{tab:pr} presents the impact of varying the poison ratio in the watermark training set. The results indicate that increasing the proportion of poisoned samples leads to a rapid improvement in VR performance. In particular, a poison ratio as low as 5\% is sufficient for \sys to achieve VR that exceeds 95\% on the NQ dataset, demonstrating the efficiency and effectiveness of the proposed watermarking scheme with minimal data modification.

\begin{table}[t]
\centering
\renewcommand{\arraystretch}{0.9}
\setlength{\tabcolsep}{3mm}
\caption{Impact of watermark training epochs on VR. Different epochs all bring high VRs exceeding 95\%.}
\label{tab:epochs}
\begin{tabular}{ccccccc}
\toprule
\textbf{Epochs} & \textbf{1}     & \textbf{2}     & \textbf{3}     & \textbf{4}     & \textbf{5}     & \textbf{10}    \\ \midrule
VR     & 97.76 & 97.19 & 97.23 & 97.76 & 98.57 & 98.90 \\ \bottomrule
\end{tabular}%
\vskip -0.15in
\end{table}
\textbf{Watermark Training Epochs.}
Table~\ref{tab:epochs} presents the influence of varying the number of watermark training epochs on VR. The results demonstrate that even a minimal number of training epochs is sufficient to achieve a high VR, consistently exceeding 95\%. This indicates the efficiency and rapid convergence of the watermark embedding process.

\subsection{Time Efficiency Analysis}
\begin{figure}[t]
    \centering
\includegraphics[width=1.0\columnwidth]{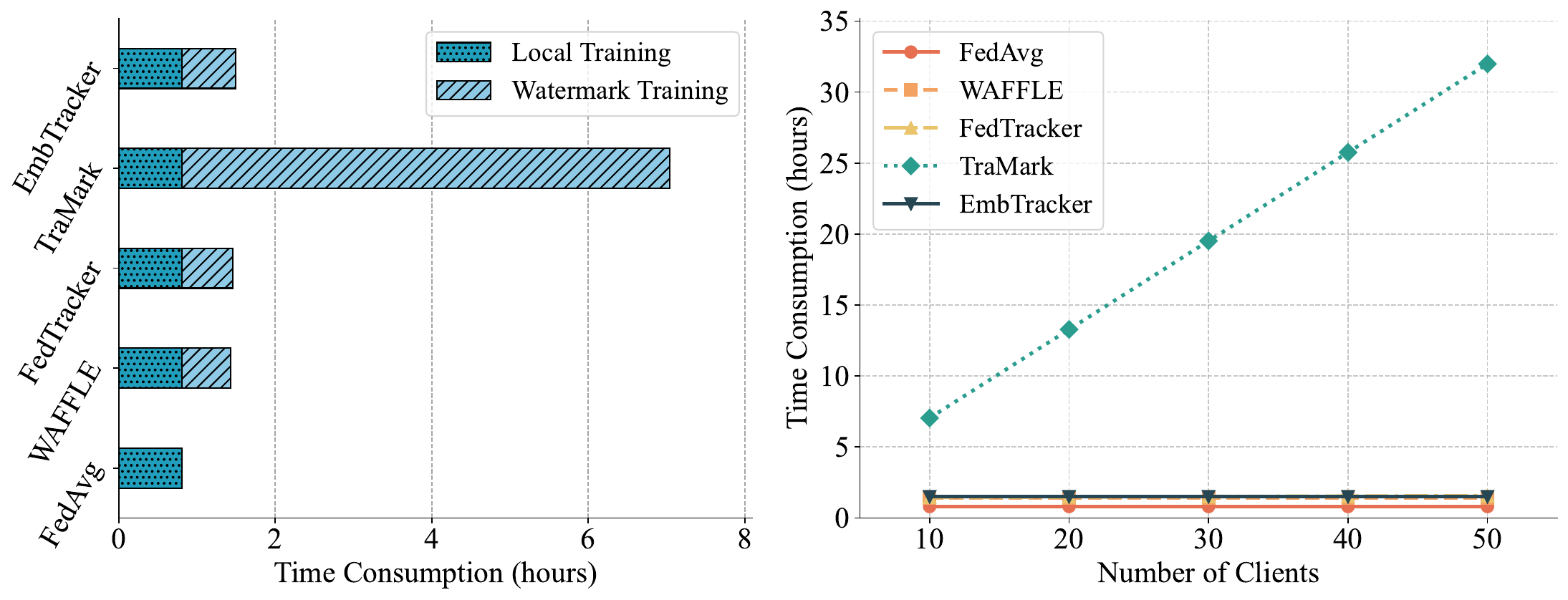}
\vspace{-18pt}
    \caption{A comparative analysis of the time overhead introduced by various watermarking schemes. All values represent the cumulative training time required for 20 rounds. The left subfigure corresponding to a scenario involving 10 clients. \sys achieves traceable black-box watermarking with minimal computational cost. Notably, \sys becomes more advantageous as the number of clients grows.}
    \label{fig:time}
    \vskip -0.2in
\end{figure}
For practical deployment considerations, we perform a systematic analysis of the time efficiency of all methods. \textbf{FedAvg} serves as the baseline, representing the workflow without watermark intervention. 
\textbf{WAFFLE} augments this process with a server-side watermark injection through additional training. This incurs a moderate increase in server-side computation each round.
\textbf{FedTracker} further incorporates a white-box fingerprint injection process that requires additional computations on each client model. The overhead introduced by FedTracker exists in every round.
\textbf{TraMark} needs substantial server-side computation to train multiple watermarked models for each client, especially as the client population increases. 

\textbf{The proposed \sys framework is designed to minimize server overhead while supporting black-box traceable watermarking}. First, the server obtains the watermark embedding using only a single time of training. Then in each round, although \sys generates a different local model for each client, it avoids training per-client watermark employing an efficient embedding replacement mechanism. The server-side watermark reinforcement steps are lightweight and do not substantially impact overall training efficiency. The client-side workflow remains unchanged and does not require additional computational or communication costs. 

As illustrated in Figure~\ref{fig:time}, empirical results demonstrate that WAFFLE and FedTracker incur moderate overhead due to their respective training and fingerprint extraction steps, while TraMark exhibits the highest time cost due to its repeated per-client retraining. In contrast, \sys introduces only a marginal increase in total training time, attributable to its simplified poisoning and replacement procedures. The overhead analysis establishes that \sys achieves a favorable balance between watermarking efficacy and computational efficiency, outperforming existing black-box traceable watermarking schemes in terms of scalability and practical deployability, and has greater advantages when more clients participate.\looseness=-1

\subsection{Robustness}
We further evaluate the robustness of \sys against its own influencing factors and possible attacks.

\begin{figure}[!t]
    \centering
\includegraphics[width=1.0\columnwidth]{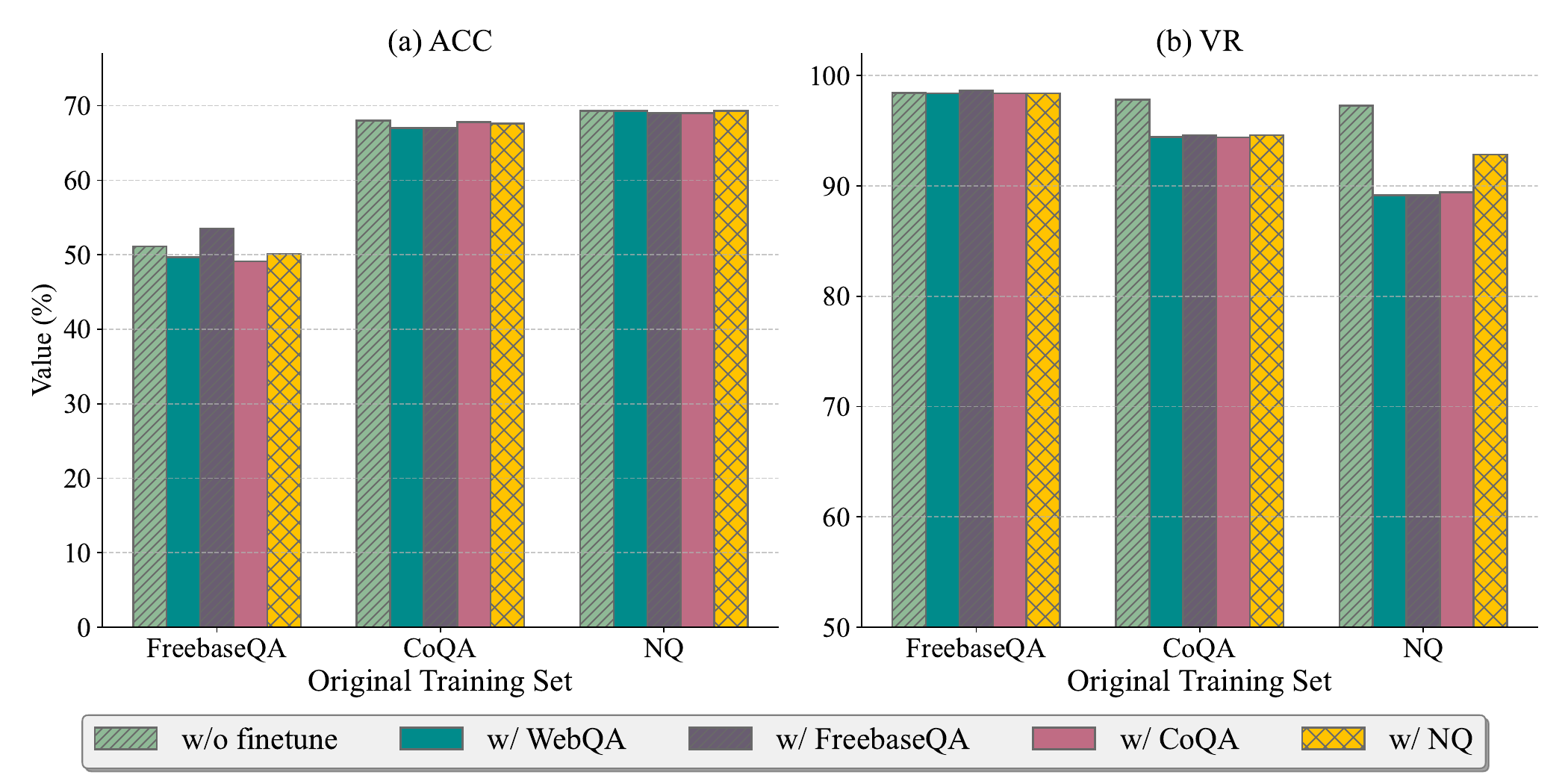}
    \vspace{-18pt}
    \caption{Robustness of \sys against fine-tuning attacks. The experiments use three client models trained on the main task and four datasets for fine-tuning attacks. Results demonstrate that the embedded watermark retains high traceability while the main task accuracy is largely preserved.\looseness=-1}
    \label{fig:finetune}
    \vskip -0.1in
\end{figure}

\textbf{Fine-tuning Attack.} To evaluate the robustness of \sys against fine-tuning attacks, we simulate malicious clients to perform additional rounds of local fine-tuning on the watermarked model using private data. Figure~\ref{fig:finetune} presents the performance before and after fine-tuning on different datasets. The results indicate that in most cases there are only minor changes in the ACC. When the main tasks are CoQA and NQ, VR decreases slightly after fine-tuning, but in almost all cases VR is still above or close to 90\%, which is a huge gap from verification leakage of less than 10\% shown in Figure~\ref{fig:vi}. This fully ensures that there is still enough confidence to confirm the identity of the model. These findings substantiate the resilience of \sys, demonstrating its effectiveness in preserving watermark traceability and model fidelity in the presence of adversarial fine-tuning attempts.
\begin{figure}[!t]
    \centering
\includegraphics[width=1.0\columnwidth]{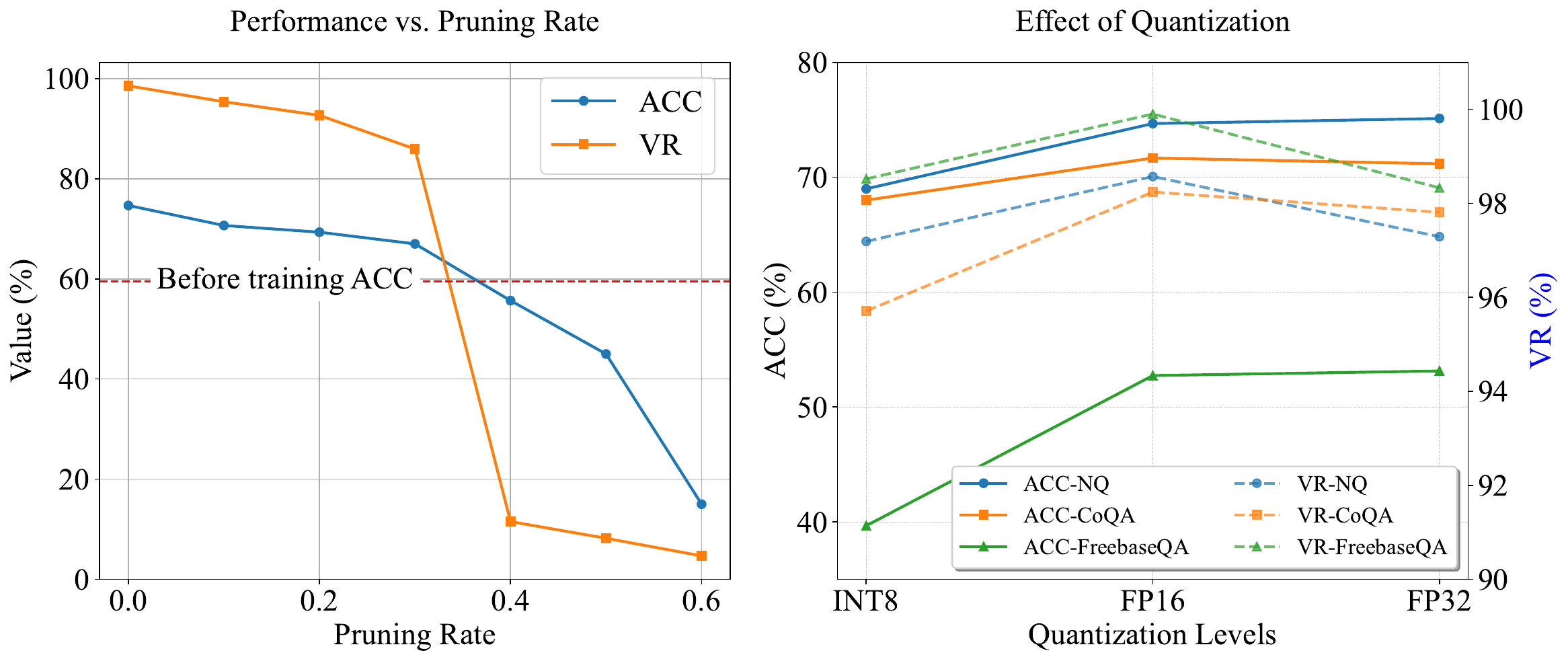}
\vspace{-18pt}
    \caption{Robustness to pruning (\texttt{Left}) and quantization (\texttt{Right}) attacks. The left sub-figure shows that \sys maintains good traceability and minimal ACC drop when no more than 30\% of the model parameters are pruned; a higher pruning rate will render the training process ineffective. The right sub-figure presents the impact of quantization levels, illustrating the resilience of the embedded watermark to quantization-induced perturbations.}
    \label{fig:prun+quantization}
    \vskip -0.15in
\end{figure}

\textbf{Pruning Attack.}
To evaluate the robustness of \sys against pruning attacks, we test different pruning rates up to 60\%. Figure~\ref{fig:prun+quantization} shows that \sys consistently maintains high VRs and negligible degradation of ACCs when up to 30\% of the model parameters are pruned (set to zero). Beyond this pruning threshold, the ACC drops lower than before training, at which point the model has become invalid and no further protection is needed. These results show that \sys is robust to moderate model pruning.\looseness=-1

\textbf{Quantization Attack.}
Due to the large memory requirements, quantization has become an important method for efficiently deploying LLMs. We further evaluate the resistance of \sys to quantization attacks under varying quantization levels. The main results in Table~\ref{tab:main} are obtained by testing under FP16 when we use FlashAttention2~\cite{dao2023flashattention2} to speed. We tested the performance under two other commonly used precisions. The right subfigure of Figure~\ref{fig:prun+quantization} shows that \sys retains a high watermark VR over 95\% although the ACC has decreased significantly. This indicates that the embedded watermark remains robust against quantization-induced perturbations, ensuring reliable ownership verification even after aggressive compression. This highlights its practicality for deployment in resource-constrained environments.

\begin{figure}[!t]
    \centering
\includegraphics[width=1.0\columnwidth]{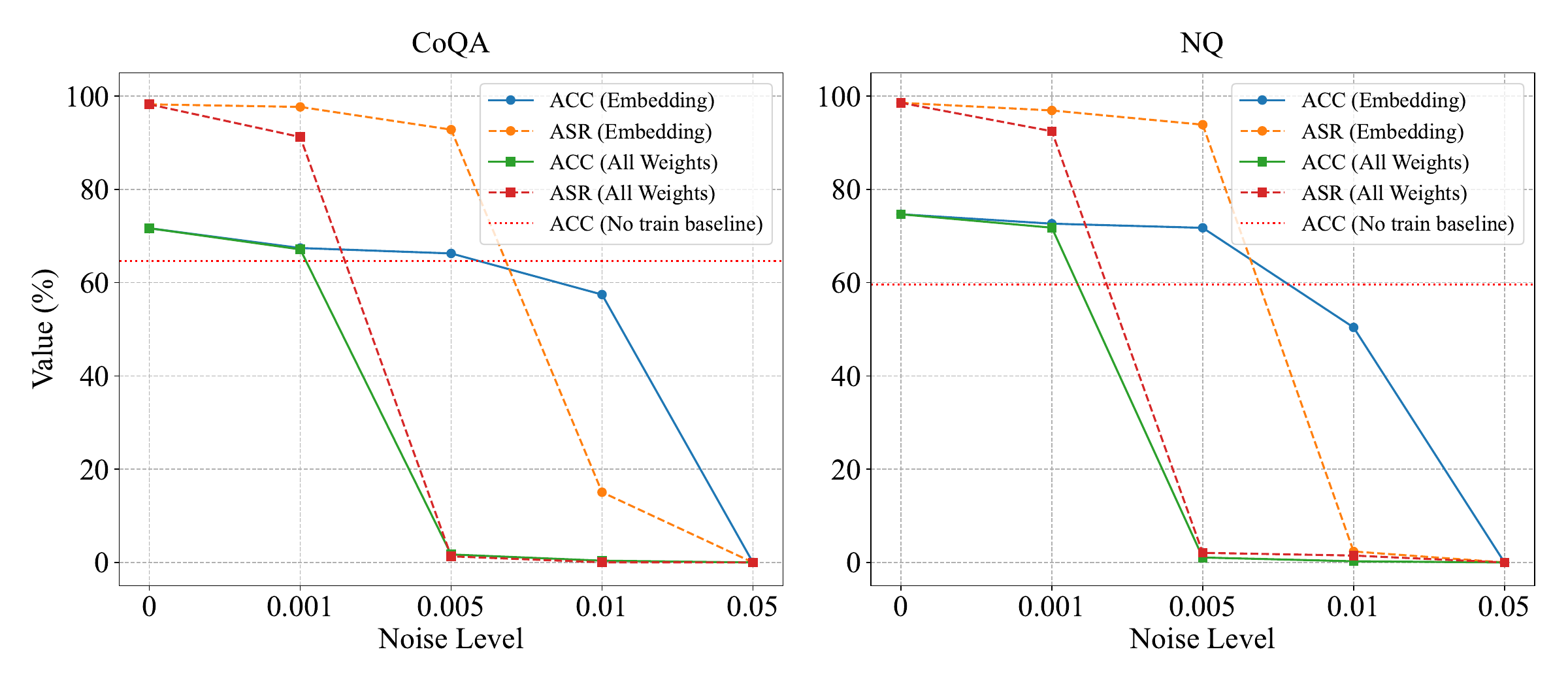}
\vspace{-18pt}
    \caption{Evaluation of noise attacks. \sys shows resilience in maintaining performance stability under varying noise interference.}
    \label{fig:noise}
    \vskip -0.15in
\end{figure}
\textbf{Noise Attack.} Figure~\ref{fig:noise} presents an analysis of the model's robustness when subjected to noise attacks. The introduction of noise can lead to a degradation in model accuracy and watermark stability, which depends on the magnitude of the noise. The watermark can still be effectively verified even when the model's performance degrades to what it was before training.
The results demonstrate the capacity of \sys to withstand perturbations and maintain performance integrity.

\subsection{Adaptive Attack}

\begin{table}[t]
\centering
\renewcommand{\arraystretch}{0.9}
\caption{Impact of overwriting attack on VR. A client is sampled as malicious attacker to perform experiments.}
\label{tab:overwrite}
\begin{tabular}{cccc}
\toprule
\multirow{2}{*}{\textbf{Setting}} & \multirow{2}{*}{\textbf{Before Attack}} & \multicolumn{2}{c}{\textbf{After Attack}}                                   \\ \cmidrule(l){3-4} 
                         &                                & \multicolumn{1}{l}{\textbf{Original Watermark}} & \textbf{New Watermark}             \\ \midrule
VR                       & 100.00                          & 98.43                                  & \multicolumn{1}{c}{99.67} \\ \bottomrule
\end{tabular}%
\vskip -0.1in
\end{table}
We further consider a more sophisticated scenario where a malicious client acts as an adaptive adversary. In this attack, we assume the adversary understands the general mechanism and attempts to disrupt the embedding vector through overwrite attacks. Overwriting means that the malicious client embeds its own watermark into the watermarked model in the same way. The adversary is unaware of the assigned embedding vector and aims to obfuscate the original embedding vector to invalidate it. We evaluate this process and test the original and new watermark as shown in Table~\ref{tab:overwrite}. It shows that after overwriting the watermark, the VR of the original watermark is still very high, which means that \sys has good robustness against the overwriting attack. Due to the sufficient watermark capacity of the LLM, the newly implanted watermark can also exist at the same time. We believe that adding a timestamp to the $sig$ used for initialization and the timestamp in the process of filing watermark information with the CA can solve the problem of the order of multiple watermarks existing simultaneously in the model.

To resolve this ambiguity, \sys can be extended with a straightforward mitigation: timestamping. \ding{182} During the Trigger Generation phase (Section 4.2), a timestamp should be cryptographically included in the identity message $m$ used to generate the digital signature $Sig$. \ding{183} When the server and clients file their watermark information with the trusted CA, this timestamp is registered alongside the triggers. In an ownership dispute where a model responds to multiple triggers, the CA can authoritatively identify the original owner by validating the earliest registered timestamp, effectively resolving the ambiguity created by the overwriting attack.

%% file: sections/7relatedworks.tex
\section{Related Work}
\textbf{Watermarking Schemes for Language Models.}
With the advent of large-scale pre-training in natural language processing, there has been more research focused on watermarking techniques tailored to LMs. RIGA~\cite{wang2021riga} introduced an auxiliary neural network to facilitate watermark embedding by utilizing weights transferred from the main network. Similarly, \cite{wu2022watermarking} presented a task-agnostic embedding loss function, yet did not explicitly address the necessity for triggers to encode model owner identity. SSLGuard~\cite{cong2022sslguard} proposed a black-box watermarking scheme for pre-trained language models, though its practical applicability is constrained by the discrete nature of word tokens. PLMmark~\cite{li2023plmmark} leveraged contrastive loss to embed backdoors in pre-trained models, enabling black-box verification on downstream classification tasks.
Hufu~\cite{xu2024hufu} proposed a modality-agnostic watermarking approach for pre-trained transformer models by exploiting the permutation equivariance property. VLA-Mark~\cite{liu2025vla} has been proposed as a cross-modal framework that embeds watermarks while preserving semantic fidelity by coordinating with vision-alignment metrics.
The Explanation as a Watermark (EaaW) method~\cite{shao2024explanation} addressed the inherent limitations of traditional backdoor-based watermarking by embedding multi-bit watermarks into feature attributions, utilizing explainable AI techniques.
Several studies~\cite{shetty2024wet,peng2023you,shetty2024warden} have explored the use of Explanation as a Service (EaaS) watermarks to protect the IP of EaaS providers. Other works~\cite{zhang2023red,shen2021backdoor} have proposed task-agnostic backdoor attacks by assigning high-dimensional vectors as trigger set labels; however, the effectiveness of these methods is often sensitive to the initialization of downstream classifiers.
Additionally, \cite{chen2025contrasting}~introduced a framework based on sets of harmless input perturbations, suggesting their utility for model fingerprinting. These approaches highlight the importance of LM watermarking, while also highlighting the ongoing challenges of achieving robust and verifiable watermarks in real-world scenarios.

\textbf{Model Watermark in Federated Learning.}
In FL, previous work has verified the feasibility of using watermarks for IP protection of models. WAFFLE~\cite{tekgul2021waffle} was the first approach to watermark DNN models in FL using additional watermark training on the server side. However, the unified watermark determines that this method is unable to identify which client leaked the model. PersistVerify~\cite{nie2024persistverifty} also embedded the unified watermark in the FL model with spatial attention and boundary sampling to verify ownership. \textit{In some frameworks, the responsibility for watermarking is placed on the clients.} For example, FedIPR~\cite{li2022fedipr} has clients embed independent watermarks in the model, facilitating subsequent verification of their IP by each client individually. However, this approach relies on client-side injection, which is impractical when malicious clients may exist within the federation. 
FedSOV~\cite{yang2023fedsov} similarly assumed that the clients are trusted and proposed a scheme that allows the clients to embed their own ownership credentials into the global model. However, this method cannot locate the malicious actor who leaked the model.

\textit{Other works have focused on achieving traceability from the server side.} FedTracker~\cite{shao2022fedtracker} introduced white-box watermark for traceability, which greatly restricted its application scenarios.  FedCIP~\cite{liang2023fedcip} added cycle-specific watermarks to different clients and located malicious clients by taking the intersection of different watermarks. However, this makes the training process very complicated, and the designed method is too idealistic to be realized. RobWE~\cite{xu2024robwe} designs watermarking specially for personalized models in personalized FL, thus this method cannot be applied to general FL model protection scenarios. TraMark~\cite{xu2025traceable} proposed a black-box watermarking framework by assigning different training sets and corresponding target outputs to each client. It is primarily designed for classification tasks and necessitates that the number of label categories be at least as large as the number of clients, while also incurring considerable computational overhead due to the need for separate watermark training for each client.
Compared to previous work, TraMark and our \sys are the first to propose a scheme to embed and verify traceable black-box watermarks from the server side, and we are the first to consider federated LMs.

%% file: sections/8conclusion.tex
\section{Conclusion}

In this paper, we propose \sys, a novel server-side framework for traceable black-box watermarking in federated LMs. By embedding identity-specific watermarks within the word embedding space, \sys facilitates the reliable attribution of model leakage while maintaining model fidelity and allowing black-box verification. Extensive experiments demonstrate that \sys delivers strong traceability, robustness, and negligible performance degradation across a wide range of federated learning settings and tasks. Our proposed framework represents an effective and practical solution for intellectual property protection in the context of federated learning for LMs.